\documentclass[aps,pre,reprint,twocolumn,superscriptaddress,showpacs]{revtex4-2}
\usepackage{amssymb,amsmath}
\usepackage[table]{xcolor}
\usepackage{graphicx}
\usepackage{mathtools}
\usepackage[export]{adjustbox}
\usepackage{overpic}
\usepackage{nicefrac}
\usepackage{multirow}
\usepackage{lipsum} 
\usepackage[normalem]{ulem}
\usepackage{pbox}
\newcolumntype{C}[1]{>{\centering\let\newline\\\arraybackslash\hspace{0pt}}m{#1}}
\usepackage{verbatim}
\usepackage{enumitem}
\usepackage{caption}
\usepackage{subcaption}
\usepackage{float}
\usepackage{framed,color}
\usepackage{verbatim}

\usepackage{bbm}

\allowdisplaybreaks

\begin{document}

	\title{Classes of non-Gaussian random matrices: long-range eigenvalue correlations and non-ergodic extended eigenvectors}
	\author{Joseph W. Baron}
	\email{jwb96@bath.ac.uk}
	\affiliation{Department of Mathematical Sciences, University of Bath, Bath, BA2 7AY, UK}
	\affiliation{Laboratoire de Physique de l’Ecole Normale Sup\`{e}rieure, ENS, Universit\'{e} PSL, CNRS, Sorbonne Universit\'{e}, Universit\'{e} de Paris, F-75005 Paris, France}
	
	\begin{abstract}
		The remarkable universality of the eigenvalue correlation functions is perhaps one of the most salient findings in random matrix theory. Particularly for short-range separations of the eigenvalues, the correlation functions have been shown to be robust to many changes in the random matrix ensemble, and are often well-predicted by results corresponding to Gaussian random matrices in many applications. In this work, we show that, in contrast, the long-range correlations of the eigenvalues of random matrices are more sensitive. Using a path-integral approach, we identify classes of statistical deviations from the Gaussian Orthogonal random matrix Ensemble (GOE) that give rise to long-range correlations. We provide closed-form analytical expressions for the eigenvalue compressibility and two-point correlations, which ordinarily vanish for the GOE, but are non-zero here. These expressions are universal in the limit of small non-Gaussianity. We discuss how these results suggest the presence of non-ergodic eigenvectors, and we verify numerically that the eigenvector component distributions of a wide variety of non-Gaussian ensembles exhibit the associated power-law tails. We also comment on how these findings reveal the need to go beyond simple mean-field theories in disordered systems with non-Gaussian interactions.
	\end{abstract}
	
	\maketitle
	
	In many applications of random matrix theory, one seeks to evaluate the average eigenvalue density of an ensemble of random matrices. The expected leading eigenvalue yields information on phase transitions in disordered systems (e.g. in spin glasses \cite{kosterlitz1976spherical,mezard1987spin}, or inference problems \cite{zdeborova2016statistical,ros2019complex}), and can be used to understand the stability of complex dynamical systems (such as neural networks \cite{rajan2006eigenvalue,aljadeff2015transition,stern2014dynamics} or complex ecosystems \cite{allesina2015stability, may1972will}). However, amongst the most striking achievements of random matrix theory is the agreement of the eigenvalue fluctuation statistics of simple Gaussian random matrices with experimental data, particularly in nuclear physics \cite{brody1981random}. A comparison of the distribution of energy level separations in heavy nuclei with 2-point eigenvalue density correlations (or Wigner's related surmise) for the Gaussian Orthogonal Ensemble (GOE) yields a remarkable agreement \cite{haq1982fluctuation,bohigas1975level, weidenmuller2009random}.
	
	Aside from nuclear physics, 2-point correlations of eigenvalues are also of central importance in other areas of physics. For example, universal Wigner-Dyson statistics \cite{mehta2004random, mirlin2000statistics, dyson1962statistical} also describe electron energy levels in disordered metals \cite{efetov1983supersymmetry}, and energy levels in chaotic quantum systems \cite{weidenmuller2009random} (e.g. the quantum billiard \cite{bohigas1984characterization}).  A justification for why the results from simple Gaussian random matrix models agree so well with such a wide variety of data is hinted at by the extraordinary robustness of the 2-point correlations to changes in the random matrix ensemble \cite{mirlin1991universality,pastur2008bulk,erdos2010universality}, so-called universality.
	
	The character of the eigenvalue density correlations can also be used to distinguish metallic from insulating phases in the context of Anderson localisation. The extended eigenvectors of the metallic phase have eigenvalues that obey GOE Wigner-Dyson statistics, while the localised eigenvectors of the insulating phase have eigenvalues that follow Poisson statistics \cite{evers2008anderson}. However, an intermediate phase has been identified in which eigenstates extend over an extensive number of sites, but with some sites being heavily weighted, and multifractality is displayed \cite{mirlin2000multifractality}. So-called non-ergodic extended states (NEES) have been found to occur around the mobility edge of the traditional Anderson model \cite{mirlin2000statistics}, and in the Anderson model on the Bethe lattice \cite{de2014anderson}. NEES have also been observed for the adjacency matrices of sparse graphs \cite{cugliandolo2024multifractal}, power-law banded random matrices \cite{vega2019multifractality}, ER graphs in the critical regime \cite{tarzia2022fully}, Laplacians of heterogeneous random networks \cite{da2024spectral}, and the generalised Rosenzweig-Porter model \cite{venturelli2023replica, biroli2021levy}. This behaviour has been associated with a non-trivial eigenvalue compressibility $\chi$ (the ratio of the variance to the mean of the number of eigenvalues in an interval) \cite{metz2017replica,venturelli2023replica,cavagna2000index,majumdar2009index}, which deviates both from that expected from both Poissonian ($\chi \approx 1$) and GOE Wigner-Dyson statistics ($\chi = 0$) \cite{cugliandolo2024multifractal, altshuler1988repulsion}.
	
	In this letter, we use a path-integral approach to identify the classes of non-Gaussian random matrix interaction statistics that defy Gaussian universality, and give rise to deviations from the GOE theory. In particular, we show which kinds of non-Gaussianity lead to an intermediate eigenvalue compressibility $0<\chi < 1$, and thus imply the presence of NEES. We see how long-range eigenvalue density correlations are responsible for the non-trivial value of $\chi$. These are absent for the GOE, where only short-range correlations on the order of the typical eigenvalue spacing are non-vanishing \cite{verbaarschot1984replica, efetov1983supersymmetry}. We derive succinct closed-form formulae for $\chi$, the eigenvalue correlations, and also the average deviations from the Wigner semi-circle law, which apply universally to all the non-Gaussian ensembles we consider (in the limit of small non-Gaussianity). We verify numerically that indeed the non-zero compressibility is associated with power-law tails in the distribution of the eigenvector components for a range of ensembles \cite{kutlin2024anatomy,cugliandolo2024multifractal}. We also discuss how these findings signal the breakdown of the usual mean-field analysis in disordered systems with non-Gaussian interactions, requiring heterogeneities to be taken into account.
	
	\textit{Random matrix ensemble and quantities of interest}---
	We consider a general class of square symmetric random matrices $\underline{\underline{a}}$ of dimension $N$, whose elements have mean $\langle a_{ij}\rangle = 0$, and possess the following moments and correlations (for $i$, $j$ and $k$ all taking different values) 
	\begin{align}
		\langle a_{ij}^2 \rangle = \frac{\sigma^2}{N} &, \hspace{1cm} \langle a_{ij}^2 a_{ik}^2 \rangle - \langle a_{ij}^2 \rangle \langle a_{ik}^2 \rangle  = \frac{\alpha_\mathrm{het} \sigma^4}{N^2} , \nonumber \\
		\langle a_{ij}^4 \rangle = \frac{\alpha_4 \sigma^4}{N} &, \hspace{1cm} 
		\langle a_{ij}a_{jk}a_{ki} \rangle = \frac{\alpha_\mathrm{cyc} \sigma^3}{N^2}, \label{stats}
	\end{align}
	where $\langle \cdots\rangle$ denotes an average with respect to realisations of the random matrix, and the scaling with $N$ ensures a sensible thermodynamic limit $N\to \infty$ \cite{mezard1987spin}. These are the most general higher-order statistics that ought to characterise the first-order deviation from the GOE (as we argue further in Supplemental Material (SM) Section S3). We neglect statistics like $\langle a_{ij} \rangle$ and $\langle a_{ij}a_{ik}\rangle$, which serve only to produce and change the location of a single outlier eigenvalue of $\underline{\underline{a}}$  \cite{benaych2011eigenvalues,edwards1976eigenvalue,baron2022eigenvalues}. Similar to the GOE, we also assume that the diagonal entries have variance $\langle a_{ii}^2 \rangle \sim 1/N$, and so are negligible for our purposes. That is, we concern ourselves only with how the statistics of the off-diagonal elements can affect the spectral behaviour. 
	
	We assume that $\alpha_4 \sim \alpha_\mathrm{het} \sim \alpha_\mathrm{cyc} \sim \alpha$, where $\alpha$ is a small parameter. In the following, we will perform a perturbative analysis to first order in $\alpha$ and $1/N$, so we assume that all higher-order correlations and moments of the matrix elements $a_{ij}$ are of the order $O(\alpha^2)$, and are therefore negligible. If we were to set $\alpha = 0$, and draw the matrix elements $a_{ij}$ from a Gaussian distribution, we would obtain a random matrix drawn from the GOE. We note that spectral statistics of $\underline{\underline{a}}$ would also be the same as the GOE ensemble in the limit $N \to \infty$ as long as the higher order moments scaled sufficiently quickly with $1/N$ \cite{erdHos2012bulk,tau2011wigner,aggarwal2019bulk,kobayakawa1995universal}, so that $\alpha \to 0$ as $N\to \infty$.
	
	Many quite natural random matrix ensembles fall into the classes defined by non-zero values of the $\alpha$-parameters. For example, the adjacency matrices of sparse Erd\H{o}s-R\'enyi (ER) graphs have $\alpha_4 \neq 0$ \cite{baron2023pathintegralapproachsparse}, as do fully-populated matrices with matrix elements drawn from heavy-tailed distributions \cite{azaele2024generalized, baron2023pathintegralapproachsparse}. The adjacency matrices of random networks with degree heterogeneity \cite{baron2022networks} (such as those constructed via the configuration or Chung-Lu models \cite{newman2018networks,chung2002connected}) and matrices with heterogeneous or hierarchical statistics \cite{poley2024eigenvalue, aljadeff2015transition, kuczala2016eigenvalue, martinez2024stabilization, giral2024interplay} have $\alpha_\mathrm{het} \neq 0$. Finally, the adjacency matrices of sparse random graphs with triangular loops \cite{metz2011spectra, pham2024effects, lopez2021transitions} and dense matrices with cyclic correlations \cite{aceituno2019universal} have $\alpha_\mathrm{cyc} \neq 0$. It is straightforward to produce hybrid ensembles for which the ratios of the $\alpha$-parameters take arbitrary values.
	
	We are interested in the statistical properties of the eigenvalues $\{\lambda_\nu\}$ of matrices $\underline{\underline{a}}$, where $\lambda_1<\lambda_2<\cdots <\lambda_N$ are real due to the symmetry of $\underline{\underline{a}}$. In particular, we consider the mean eigenvalue density and the eigenvalue correlations (respectively)
	\begin{align}
		\rho(\omega) =& \left\langle\frac{1}{N} \sum_{\nu} \delta\left(\omega - \lambda_\nu \right) \right\rangle \nonumber \\
		\rho_c(\omega, \mu) =& \frac{\sum_{\nu, \nu'} \left\langle[\delta(\omega - \lambda_{\nu})-\rho(\omega)][\delta(\mu - \lambda_{\nu'}) -\rho(\mu)] \right\rangle}{N^2} ,
	\end{align}
	where $\delta(\cdot)$ is the Dirac delta function. We can also consider the statistics of the number of eigenvalues in an interval $\mathcal{I}_N(E,s) = \int_{E-s/2}^{E+s/2} d\omega \sum_\nu \delta(\omega - \lambda_\nu)$. The eigenvalue compressibility \cite{metz2017replica,venturelli2023replica,cavagna2000index,majumdar2009index}, which is defined as the ratio of the variance of $\mathcal{I}_N(E,s)$ to its mean, is related to the eigenvalue density and correlations via 
	\begin{align}
		\chi(E,s) = \frac{N\int_{E-s/2}^{E+s/2} d\omega\int_{E-s/2}^{E+s/2}  d\mu \rho_c(\omega,\mu) }{\int_{E-s/2}^{E+s/2} d\omega \rho(\omega)} . \label{compressibilitydef}
	\end{align}
	The eigenvalue compressibility acts as an indicator of the localisation properties of the eigenvectors in the system. When eigenvectors are localised, such that only a small number $O(N^0)$ of their components are non-zero, the eigenvalues obey Poisson statistics and $\chi \sim 1$. When the eigenvectors are extended, as they are for the GOE with all $N$ components being of similar magnitude, one has $\chi \to 0$ for $N\to \infty$ \cite{cugliandolo2024multifractal, altshuler1988repulsion}. 
	
	\begin{figure*}[t]
		\centering 
		\includegraphics[scale = 0.25]{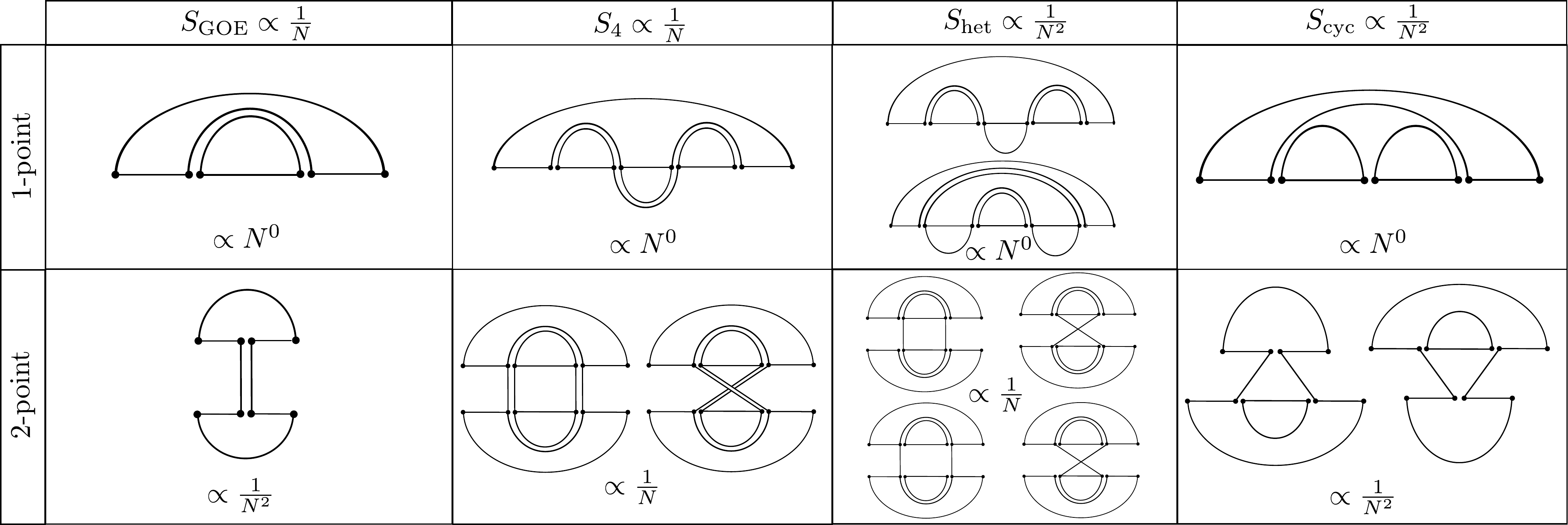}
		\captionsetup{justification=raggedright,singlelinecheck=false, font=small}
		\caption{Some example Feynman diagrams that contribute to the series expansions of the 1- and 2-point Green's functions. The various classes of non-Gaussianity (non-trivial fourth moment, statistical heterogeneity, cyclic correlations) give rise to diagrams with differing topologies. The order of the diagram in $1/N$ (indicated above for each case) is determined by both the topology and the order of the diagram in $S_\mathrm{GOE}$, $S_4$, $S_\mathrm{het}$ or $S_\mathrm{cyc}$ (all pictured diagrams are first-order). Details on how these diagrams are derived and how they can be used to calculate the Green's functions are given in the SM. }\label{fig:feynman}
	\end{figure*}
	
	\textit{GOE-like matrices}---
	To contrast against later results, we briefly recall the corresponding well-known properties of GOE matrices (and those within the GOE universality class) for $N \to \infty$. Most famously, the expected density of eigenvalues for the GOE tends to the celebrated Wigner semi-circle law \cite{wigner1958distribution, wigner1967random}. Importantly for the present discussion, the connected two-point correlations of the eigenvalues follow \cite{verbaarschot1984replica} 
	\begin{align}
		\rho_c(\omega, \mu)   \sim& -[N(\omega-\mu)]^{-2}, \label{goemacrotwopoint}
	\end{align}
	for long-range eigenvalue separations $\vert \mu - \omega\vert \gg N^{-1}$. For short ranges  $\vert \mu - \omega\vert \ll N^{-1}$, the two-point correlations instead obey an integral expression involving a sine kernel \cite{efetov1983supersymmetry}. To probe the short-range statistics, one can also examine the nearest-neighbour eigenvalue spacings, which obey the Wigner surmise \cite{weidenmuller2009random, gaudin1961loi} (see also SM Section S3.2). One has for the eigenvalue compressibility in the GOE case \cite{metz2017replica,venturelli2023replica,cavagna2000index,majumdar2009index}
	\begin{align}
		\chi  \sim \ln( N)/N  , \label{compressibilitygoe}
	\end{align}
	so that $\chi \to 0$ as $N\to \infty$. Finally, the eigenvector components $\psi_i^{(\nu)}$, corresponding to the eigenvalue $\lambda_\nu$, are known to be independent centred Gaussian random variables with variance $N^{-1}$ \cite{jiang2006how} (under the normalisation $\sum_i \vert \psi_i^{(\nu)}\vert^2 = 1 $). 
	
	\textit{Perturbative treatment about the GOE case and Feynman diagrams}---
	We now calculate the eigenvalue spectral density and the 2-point correlations of the eigenvalue fluctuations for the non-Gaussian ensembles with statistics in Eq.~(\ref{stats}). As is common for calculations in random matrix theory, we proceed via the intermediate step of calculating the 1- and 2-point matrix Green's functions $G(\omega) = N^{-1}\sum_i \langle g_i(\omega) \rangle$ and $G_c(\omega, \mu) = N^{-2}\sum_{ij} \langle[g_i(\omega)-G(\omega)][g_j(\mu)-G(\mu)]\rangle$, where $g_i(\omega) = ([\omega\underline{\underline{\mathbbm{1}}}- \underline{\underline{a}}]^{-1})_{ii}$. Following Ref. \cite{baron2022eigenvalues}, we do this by considering a particular linear dynamical system [see SM Section S1], for which the matrix $\underline{\underline{a}}$ quantifies the interactions between components. We then use the Martin-Siggia-Rose-Janssen-de Dominicis (MSRJD) path-integral formalism \cite{altlandsimons, hertz2016path} to evaluate the Green's functions. The advantage of using the path-integral representation is that it facilitates a perturbative analysis using Feynman diagrams. 
	
	The main challenge is thus the evaluation of the Green's functions from their path-integral expressions. We can show in general that the MSRJD action takes the following form when the matrix elements have the statistics in Eq.~(\ref{stats})
	\begin{align}
		S = S_0 + S_\mathrm{GOE} + \alpha_4 S_\mathrm{4} + \alpha_\mathrm{het} S_\mathrm{het} + \alpha_\mathrm{cyc} S_\mathrm{cyc} + O(\alpha^2) . \label{goepluspert}
	\end{align}
	Explicit expressions for $S_0$, $S_\mathrm{GOE}$, $S_\mathrm{4}$, $S_\mathrm{het}$ and $S_\mathrm{cyc}$ are given in the SM [see Eq.~(S46)]. Assuming that $\alpha \ll 1$, we perform a diagrammatic expansion of the path-integral expression for the Green's functions, keeping only contributions that are leading order in $1/N$, and also first order in the perturbation parameters $\alpha$. We identify Feynman diagrams corresponding to each class of non-Gaussianity, as shown in Fig. \ref{fig:feynman}. 
	
	\newpage
	
	\textit{Long-range correlations, non-zero compressibility and NEES}---~Upon resumming the appropriate diagrammatic series in  the thermodynamic limit, we obtain expressions for the spectral observables valid to leading order in $\alpha$. First, we find that the eigenvalue density is close to the Wigner semi-circle law, with deviations of the order $O(\alpha)$, which we quantify. This echoes the result of Rodgers and Bray \cite{rodgers1988density}, who examined the case of sparse ER graphs. The general expression for $\rho(\omega)$ is given in SM Section S4, where it is checked against numerics for a range of random matrix ensembles. 
	
	In contrast to the eigenvalue density, the 2-point functions are altered more drastically by the non-Gaussianity. In fact, the scaling of $\rho_c(\omega, \mu)$ with $N$ is different to the GOE case in Eq.~(\ref{goemacrotwopoint}). This is found immediately by studying the 2-point Feynman diagrams associated with the perturbations (see Fig. \ref{fig:feynman}). Upon resumming the appropriate diagrammatic series to leading order in $1/N$, one finds [see SM Section S5]
	\begin{align}
		\rho_c(\omega, \mu) =& \frac{\overline{\alpha}_2}{4\pi^2 N} f(\omega) f(\mu) + O\left(\frac{1}{N^2}\right)+ O\left(\frac{\alpha^2}{N}\right),  \label{2pointsparsehet}
	\end{align}
	where $f(\cdot)$ is an elementary expression, which is given in the SM. The expression in Eq.~(\ref{2pointsparsehet}) is verified directly in Fig. S3 of the SM. Importantly, we see that the eigenvalue correlations are proportional to $\overline{\alpha}_2 = 2 \alpha_4 + 4 \alpha_\mathrm{het}$. We note that there is no contribution from $\alpha_\mathrm{cyc}$ in $\overline{\alpha}_2$ because the 2-point Feynman diagrams that arise from $S_\mathrm{cyc}$ are $\propto N^{-2}$, like those of the GOE (see Fig. \ref{fig:feynman}). 
	
\begin{figure}[t]
		\centering 
		\includegraphics[scale = 0.5]{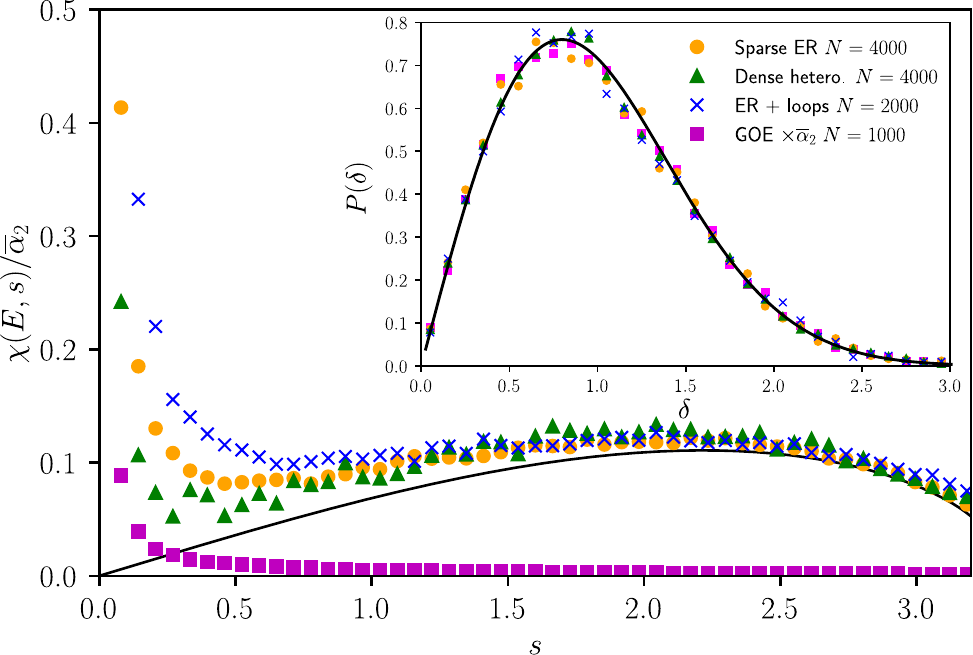}
		\captionsetup{justification=raggedright,singlelinecheck=false, font=small}
		\caption{Verification of the universal formula for the eigenvalue compressibility in Eq.~(\ref{chisparsehet}) (solid black line). GOE results are given for comparison. Deviations from the theory are due to eigenvalue repulsion at small separations, and they vanish as $N\to \infty$. We consider various random matrix ensembles, each with different values of $\alpha_4$, $\alpha_\mathrm{het}$ and $\alpha_\mathrm{cyc}$, which are described in detail in SM Section S3.2. Inset: Comparison of nearest neighbour eigenvalue separations for the same ensembles to the Wigner surmise (also given in SM Section S3.2).  }\label{fig:chi}
	\end{figure}

	One notes that while the perturbative approach provides us with information about the long-range eigenvalue correlations, the short-range correlations are a non-perturbative contribution to $\rho_c(\omega, \mu)$ in $1/N$ \cite{brezin1994correlation}, and are therefore better-treated using methods such as orthogonal polynomials \cite{brezin1993universality} or the supersymmetric approach \cite{verbaarschot1985critique,mirlin1991universality,efetov1983supersymmetry}. However, as was shown analytically in Ref. \cite{mirlin1991universality} specifically for sparse random matrices, we also observe here that the short-range eigenvalue statistics are not modified for the non-Gaussian ensembles (see inset of Fig. \ref{fig:chi}); only the long-range eigenvalue correlations are affected substantially.
	
	Given the expression for the eigenvalue density correlations in Eq.~(\ref{2pointsparsehet}), it is straightforward to obtain the eigenvalue compressibility, as defined in Eq.~(\ref{compressibilitydef}). One finds
	\begin{align}
		\chi(E, s) = \frac{\overline{\alpha}_2}{4\pi^2} \frac{F^2(E,s)}{R(E,s)} + O\left(\log(N)/N\right) + O\left(\alpha^2\right) , \label{chisparsehet}
	\end{align}
	where $F(E,s) = \int_{E-s/2}^{E+s/2}d\omega f(\omega)$ and $R(E,s) = \int_{E-s/2}^{E+s/2}d\omega \rho(\omega)$. Given the elementary expressions for $\rho(\omega)$ and $f(\omega)$ that have been obtained, the explicit computation of $F(E,s)$ and $R(E,s)$ is straightforward. We note that the form of $\chi(E,s)$ is universal for all of the non-Gaussian ensembles that we consider; the particulars of the ensemble only enter via the constant $\bar\alpha_2$.

\begin{figure}[t]
		\centering 
		\includegraphics[scale = 0.5]{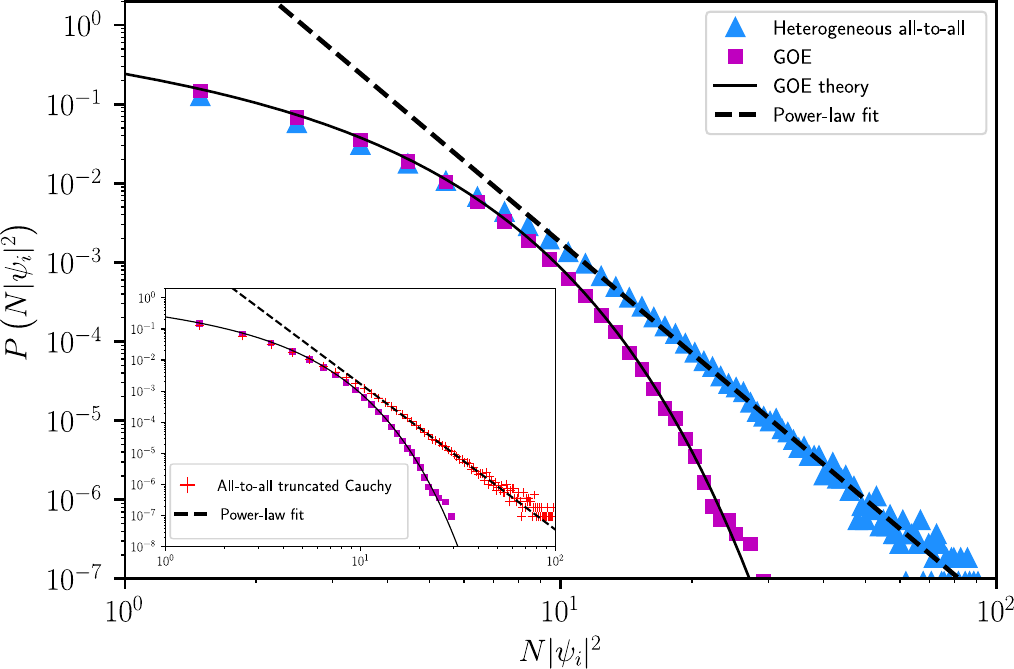}
		\captionsetup{justification=raggedright,singlelinecheck=false, font=small}
		\caption{ Distribution of the squared eigenvector components for fixed eigenvalue $\omega = 0.4$.  The random matrix ensembles with $0<\chi<1$ exhibit deviations from the GOE result in the form of power-law tails. Letting $y = N\vert\psi_i\vert^2$, the solid line is the GOE result $P = \frac{1}{\sqrt{2\pi y}} e^{-y/2}$. The dashed line is a power-law fit $P = \beta y^\gamma$ (with $\beta$ and $\gamma$ constants) to the data from an ensemble with $\alpha_\mathrm{het} \neq 0$ (see SM S3.2 for details). Inset: A similar plot for an ensemble with $\alpha_4\neq 0$. This same behaviour is verified for other ensembles in the SM (see Fig. S5). }\label{fig:vectors}
	\end{figure}

	Crucially, we see that the eigenvalue compressibility has become non-vanishing in the thermodynamic limit. This is verified for a range of random matrix ensembles in Fig. \ref{fig:chi}. That the eigenvalue compressibility is non-vanishing, but differs from what would be expected by Poisson statistics, is known to occur in sparse graphs \cite{metz2016large, cugliandolo2024multifractal, metz2015index} (which have $\alpha_4 \neq 0$), in agreement with the present results. This behaviour has been linked to the emergence of non-ergodic extended states (NEES) and multifractal eigenvectors \cite{biroli2012difference,de2014anderson}, which have also been observed in heterogeneous graphs \cite{da2024spectral} (which have $\alpha_\mathrm{het} \neq 0$). From the result in Eq.~(\ref{chisparsehet}), we expect that NEES should be seen more generally in ensembles with non-zero $\alpha_4$ or $\alpha_\mathrm{het}$. Indeed, we verify the presence of power-law tails in the distribution of eigenvectors, characteristic of NEES \cite{cugliandolo2024multifractal, kutlin2024anatomy}, for a range of ensembles with $\alpha_\mathrm{het},\alpha_{4} \neq 0$ in Fig. \ref{fig:vectors} and SM Fig. S5. 
	
	\textit{Further implications}---
	The results presented here have significance in addition to NEES. The scaling of $\chi(E,s)$ with $N$ also implies a particular scaling of the large deviation function for the typical number of eigenvalues in an interval. Letting $\mathcal{I}_N = \int_0^\infty d\omega \sum_\nu \delta(\omega - \lambda_\nu)$,  the probability distribution $\mathcal{P}(c, N) = \mathrm{Prob}(\mathcal{I}_N = c N)$ takes the following form in the GOE case $\mathcal{P}(c, N) \sim \exp[- \pi^2 N^2 (c-1/2 )^2/(2 \ln N)]$ \cite{cavagna2000index,majumdar2009index}. One thus recovers the scaling in Eq.~(\ref{compressibilitygoe}) by computing $\chi = (\langle \mathcal{I}_N^2 \rangle - \langle\mathcal{I}_N\rangle^2)/\langle \mathcal{I}_N\rangle$. For sparse random matrices and L\'evy random matrices (both with $\alpha_4\neq 0$), it has been shown instead that $\mathcal{P}(c, N) \sim \exp[- N \Phi(c)]$ \cite{metz2015index, metz2016large, bordenave2024large}. The result in Eq.~(\ref{chisparsehet}) is in keeping with these works, and it would also suggest that the large deviation function for random matrix ensembles with $\alpha_\mathrm{het}\neq 0$ ought also to have the same scaling. 
	
	It is also possible to study other quantities using the diagrammatic approach (see SM Section S5.2) such as the correlations of the local density of states $\rho_i(\omega) = \lim_{ \epsilon\to 0}\mathrm{Im} \left\{([(\omega-i \epsilon) \underline{\underline{\mathbbm{1}}} - \underline{\underline{a}} ]^{-1})_{ii}\right\}$, defined as $K_c(\omega, \mu) = N^{-1}\sum_{i} \langle [\rho_i(\omega)-\rho(\omega)] [\rho_i(\mu)-\rho(\mu)] \rangle$, which is closely related to the overlap correlation function \cite{mirlin2000statistics, cugliandolo2024multifractal}. We find that a non-zero value of $K_c$ also arises when $\alpha_4\neq 0$ or $\alpha_\mathrm{het} \neq 0$, while $K_c$ vanishes for the GOE. We demonstrate in SM Section S6 that this behaviour can be associated with the breakdown of simple mean-field theory in systems  with interaction statistics like those in Eq.~(\ref{stats}). This is in keeping with previous studies involving non-Gaussian interactions, which found that a heterogeneous mean-field theory (HMFT) with non-Gaussian noise terms becomes necessary \cite{aguirre2024heterogeneous, peralta2020binary,azaele2024generalized}. We comment on why this is true, and we replicate the diagrammatic findings for $\rho(\omega)$ and $K_c$, using a general HMFT. However, we emphasise that even the HMFT approach is inadequate when computing $\rho_c$ (and therefore $\chi$), for which one must take into account correlations beyond the mean-field. These are captured by the diagrammatic approach used above.
	
	\textit{Conclusion}---
	In this work, we have demonstrated that while the long-range eigenvalue correlations vanish for Gaussian i.i.d. random matrices, they are non-trivial when the random matrix ensemble in question deviates from Gaussian statistics in specific ways, which we have identified. This gives rise to an eigenvalue compressibility $0<\chi<1$, which has been linked with the presence of non-ergodic extended states \cite{metz2016large, cugliandolo2024multifractal, metz2015index, altshuler1988repulsion}. Indeed, we verified that the eigenvector components have power-law-tailed distributions under these circumstances. We have also shown that the results for the 2-point Green's functions derived here have implications for the scaling of the large deviation function for the number of eigenvalues in an interval \cite{metz2015index, metz2016large,bordenave2024large} and the breakdown of simple mean-field theories \cite{aguirre2024heterogeneous, peralta2020binary,azaele2024generalized}, demonstrating the broad significance of the classes of non-Gaussianity that we have identified.
	
	Although the results for the eigenvalue correlations obtained here could in principle also be obtained via a replica approach for example \cite{metz2017replica,venturelli2023replica,cavagna2000index}, one advantage of the diagrammatic approach is its generalisability. We have shown that diagrammatic series can be used to compute a variety of quantities for a broad class of matrix ensemble. While we have only considered first-order corrections to the GOE case, one could compute higher-order corrections without much difficulty, as was done in Ref. \cite{baron2023pathintegralapproachsparse} for the eigenvalue density. One could also extend the present consideration to the non-Hermitian case. Indeed, 2-point Green's functions have been computed for non-Hermitian ensembles using diagrammatic methods  \cite{janik1997non,nowak2018probing}.

	\acknowledgements
	The author would like to thank Giulio Biroli, Tobias Galla, Tim Rogers, Marco Tarzia, and Davide Venturelli for insightful and helpful discussions. This work was supported by grants from the Simons Foundation (\#454935 Giulio Biroli). The author thanks the Leverhulme Trust for support through the Leverhulme Early Career Fellowship scheme.

\end{document}


\title{Supplemental Material: \\ Long-range eigenvalue density correlations of non-Gaussian and structured random matrices}
	
	\title{Classes of non-Gaussian random matrices: long-range eigenvalue correlations and non-ergodic extended eigenvectors\\ \medskip \medskip{\Large--- Supplemental Material ---}}
	\author{Joseph W. Baron}
	\email{jwb96@bath.ac.uk}
	\affiliation{Department of Mathematical Sciences, University of Bath, Bath, BA2 7AY, UK}
	\affiliation{Laboratoire de Physique de l’Ecole Normale Sup\`{e}rieure, ENS, Universit\'{e} PSL, CNRS, Sorbonne Universit\'{e}, Universit\'{e} de Paris, F-75005 Paris, France}

	\maketitle

	\onecolumngrid
	
	\tableofcontents
	
	\pagebreak
	
	\section{General method}
	\subsection{One- and two-point response functions and their relationship with the eigenvalue spectrum}
	Consider a random matrix $\underline{\underline{a}}$. Following Ref. \cite{baron2022eigenvalues}, suppose we construct the following linear dynamical system
	\begin{align}
		\dot x_i = -\omega x_i + \sum_{j}a_{ij}x_j + h^{(x)}_i(t). \label{dynamicalsystem}
	\end{align}
	Let us define the response function as the functional derivative $R^{(x)}_{ij}(t,T) = \delta x_i(t)/\delta h^{(x)}_j(T) \vert_{h = 0}$. We see that the Laplace transforms of the response functions are related to the resolvent (or matrix Green's function) of the random matrix
	\begin{align}
		g_i(\omega) \equiv \lim_{\nu \to 0}\mathcal{L}_t\left[ R^{(x)}_{ii}(t,0)\right](\nu) =  \left[\left(\omega\underline{\underline{\id}}  - \underline{\underline{a}} \right)^{-1}\right]_{ii},
	\end{align}
	where $\mathcal{L}_t[f(t)](\eta) = \int_0^t dt f(t) e^{-\eta t}$ is the Laplace transform. Defining $G(\omega) = N^{-1} \sum_i \langle g_i(\omega)\rangle$, where $ \langle \cdots\rangle$ denotes an average over realisations of $\underline{\underline{a}}$, one can extract the average eigenvalue density via
	\begin{align}
		\rho(\omega) = \lim_{\epsilon \to 0} \mathrm{Im} G(\omega - i \epsilon)/\pi . \label{densfromres}
	\end{align}
	We can also define the connected two-point Green's function via 
	\begin{align}
		G_c(\omega, \mu) =N^{-2} \sum_{i,j} \left[\langle g_i(\omega)g_j(\mu)\rangle -  \langle g_i(\omega)\rangle\langle g_j(\mu)\rangle \right], 
	\end{align}
	from which one can obtain the eigenvalue density fluctuations using
	\begin{align}
		\rho_c(\omega, \mu)  = -\frac{1}{4 \pi^2} \lim_{\epsilon, \delta \to 0}\left[ G_c(\omega + i \epsilon, \mu + i \delta) + G_c(\omega - i \epsilon, \mu - i \delta) -  G_c(\omega + i \epsilon, \mu - i \delta) -  G_c(\omega - i \epsilon, \mu + i \delta)\right] .\label{corrfromgc}
	\end{align}
	One can also define the local density of states
	\begin{align}
		\rho_i(\omega) = \lim_{\epsilon \to 0} \mathrm{Im} g_i(\omega - i \epsilon)/\pi . \label{ldosdef}
	\end{align}
	If we define the alternative two-point Green's function 
	\begin{align}
		H_c(\omega, \mu) = N^{-1} \sum_i \langle g_i(\omega)g_i(\mu) \rangle - N^{-2} \sum_{ij} \langle g_i(\omega) \rangle \langle g_j(\mu) \rangle , \label{hcdef}
	\end{align}
	we can extract the covariance of the local density of states
	\begin{align}
	  K_c (\omega,\mu) = -\frac{1}{4 \pi^2} \lim_{\epsilon, \delta \to 0}\left[ H_c(\omega + i \epsilon, \mu + i \delta) + H_c(\omega - i \epsilon, \mu - i \delta) -  H_c(\omega + i \epsilon, \mu - i \delta) -  H_c(\omega - i \epsilon, \mu + i \delta)\right] .\label{corrfromkc}
	\end{align}		
	One can also examine the typical number of eigenvalues in an interval $\mathcal{I}_N(E,s) = \int_{E-s/2}^{E+s/2} d\omega \sum_\nu \delta(\omega - \lambda_\nu)$, and the variance of this number
	\begin{align}
		\langle \mathcal{I}_N(E,s) \rangle &= N\int_{E-s/2}^{E+s/2} d\omega \rho(\omega), \nonumber \\
		\langle [\mathcal{I}_N(E,s)]^2 \rangle - \langle \mathcal{I}_N(E,s) \rangle^2 &= N^2\int_{E-s/2}^{E+s/2} d\omega \int_{E-s/2}^{E+s/2} d\mu \rho_c(\omega,\mu).
	\end{align}
	Finally, we define the eigenvalue compressibility via
	\begin{align}
		\chi(\Delta; E) = \frac{\langle [\mathcal{I}_N(E,s)]^2 \rangle - \langle \mathcal{I}_N(E,s) \rangle^2}{\langle \mathcal{I}_N(E,s) \rangle } = N\frac{\int_{E-s/2}^{E+s/2} d\omega \int_{E-s/2}^{E+s/2} d \mu \, \rho_c(\omega, \mu)  }{\int_{E-s/2}^{E+s/2} d\omega \,\rho(\omega)} . \label{compressibilitydef}
	\end{align}
	One notes that this quantity varies between $\chi \to 0$ for GOE matrices as $N \to \infty$ (due to spectral rigidity) and $\chi \sim 1$ for Poisson statistics. More precisely, we have for $s \gg N^{-1}$ 
	\begin{align}
		\chi(E,s) \propto \frac{\ln\left[ N \rho(E)\Delta\right]}{N \rho(E)\Delta} ,
	\end{align}
	for the usual Wigner-Dyson GOE statistics \cite{mehta2004random}. In the case where the eigenvalues have no level correlations and are independently distributed (i.e. Poisson statistics), one instead has \cite{venturelli2023replica}
	\begin{align}
		\chi(E,s) = 1 - \langle I_N(E,s) \rangle .
	\end{align}

	\subsection{Path-integral construction}
	We now introduce the Martin-Siggia-Rose-Janssen-de Dominicis (MSRJD) path integral \cite{altlandsimons, msr, janssen1976lagrangean, dominicis1976techniques, dedominicis1978dynamics, hertz2016path} that will be the cornerstone of our subsequent analysis and provide us with the disorder-averaged response functions of the dynamical system in Eq.~(\ref{dynamicalsystem}). 
	
	The MSRJD path integral that we consider is the generating functional for the dynamical process in Eq.~(\ref{dynamicalsystem}) \cite{altlandsimons}. As such, the time-dependent correlators and response functions of the quantities $x_i(t)$ can be found by taking appropriate functional derivatives of this object. 
	
	For the sake of calculating the 2-point response functions, it is convenient to introduce a replicated set of dynamical variables 
	\begin{align}
		\dot x_i &= -\omega x_i + \sum_{j}a_{ij}x_j + h^{(x)}_i(t), \nonumber \\
		\dot y_i &= -\mu y_i + \sum_{j}a_{ij}y_j + h^{(y)}_i(t). \label{replicatedprocess}
	\end{align}
	Averaged over realisations of the matrix $\underline{\underline{a}}$, the functional integral for the coupled process is written
	\begin{align}
		&Z[\psi^{(x)}, h^{(x)}, \psi^{(y)}, h^{(y)}] = \int D[x, \hat x, y, \hat y] \Bigg\langle\exp\left[i \sum_{i}\int dt \psi^{(x)}_i x_i + i\sum_{i} \int dt \, \hat x_i \left(\dot x_i + \omega x_i- \sum_{ j}a_{ij} x_j - h^{(x)}_i \right)\right]\nonumber \\
		 &\times\exp\left[i \sum_{i}\int dt \psi^{(y)}_i y_i + i\sum_{i} \int dt \, \hat y_i \left(\dot y_i + \mu y_i- \sum_{ j}a_{ij} y_j - h^{(y)}_i \right)\right]\Bigg\rangle, \label{genfunct}
	\end{align}
	where $D[x, \hat x, y, \hat y]$ indicates integration with respect to all possible trajectories of the variables $\{ x_i(t), y_i(t)\}$ and their conjugate `momenta' $\{ \hat x_i(t), \hat y_i(t) \}$, and we remind the reader that $\langle \cdots \rangle$ (without a subscript) denotes an average with respect to realisations of the random matrix entries. Constant factors that ensure the normalisation $Z[0,h^{(x)}, 0, h^{(y)}] = 1$ have been absorbed into the integral measure. Aside from the source terms containing the variables $\psi^{(x)}$ and $\psi^{(y)}$, the integrand in Eq.~(\ref{genfunct}) is merely a complex exponential representation of Dirac delta functions, which constrain the system to follow trajectories satisfying Eqs.~(\ref{replicatedprocess}), averaged over realisations of the random matrix entries. The reader is directed to Refs. \cite{altlandsimons, hertz2016path} for further details.
	
	The disorder-averaged response functions of the system can be found from this object by differentiating as follows
	\begin{align}
		\left\langle R^{(x)}_{ii}(t, T) \right\rangle &= \frac{\delta \langle x_i(t)\rangle_S}{\delta h^{(x)}_i(T)} \bigg\vert_{\psi = h = 0} = -i\frac{\delta^2 Z}{\delta \psi^{(x)}_{i}(t) \delta h^{(x)}_i(T)} \bigg\vert_{\psi = h = 0} \nonumber \\
		&= - i  \left\langle x_i(t) \hat x_i(T) \right\rangle_S \big\vert_{\psi = h = 0} , \label{responsefunctions}
	\end{align}
	where here $\langle \cdots \rangle_S$ indicates an average with respect to the dynamical process, i.e.
	\begin{align}
		&\langle \mathcal{O} \rangle_S \big\vert_{\psi = h =0}  = \int D[x, \hat x, y, \hat y] \,\,\mathcal{O} \,\,\Bigg\langle\exp\left[ i\sum_{i} \int dt \, \hat x_i \left(\dot x_i + \omega x_i - \sum_{ j}a_{ij} x_j  \right)\right]\nonumber \\
		&\times\exp\left[ i\sum_{i} \int dt \, \hat y_i \left(\dot y_i +\mu y_i- \sum_{ j}a_{ij} y_j  \right)\right]\Bigg\rangle \nonumber \\
		&\equiv \int D[x, \hat x, y, \hat y] \,\,\mathcal{O} \,\,e^S, \label{disorderaverageobservable}
	\end{align}
	where we have defined the MSRJD action $S$. From now on, it is to be understood that all averages $\langle \cdot\rangle_S$ are to be evaluated at $\psi = h = 0$. We note that averages involving only the conjugate variables evaluate to zero
	\begin{align}
		- \langle \hat x_i(t) \hat x_j(T) \rangle_S = \frac{\delta^2 Z}{\delta h^{(x)}_{i}(t) \delta h^{(x)}_i(T)} \bigg\vert_{\psi = h = 0} = \frac{\delta^2 [Z \vert_{\psi = 0}]}{\delta h^{(x)}_{i}(t) \delta h^{(x)}_i(T)} \bigg\vert_{ h = 0} = \frac{\delta^2 [1]}{\delta h^{(x)}_{i}(t) \delta h^{(x)}_i(T)} \bigg\vert_{ h = 0} = 0 . \label{hatzero}
	\end{align}

	More complicated quantities, such as the connected two-point response functions required for the 2-point eigenvalue correlations, can also be extracted. We have
	\begin{align}
		G_c(\omega, \mu) = \lim_{\nu, \xi \to 0}\mathcal{L}_{t'}\left[\mathcal{L}_t\left[ N^{-2}\sum_{ij }\left[ \langle R^{(x)}_{ii}(T,0)R^{(y)}_{jj}(T',0) \rangle - \langle R^{(x)}_{ii}(T,0)\rangle \langle R^{(y)}_{jj}(T',0) \rangle \right] \right](\nu)\right](\xi) , \label{2pointgc}
	\end{align}
	which can be written as a path integral expression using
	\begin{align}
		\langle R^{(x)}_{ii}(T,0)R^{(y)}_{jj}(T',0) \rangle - \langle R^{(x)}_{ii}(T,0)\rangle \langle R^{(y)}_{jj}(T',0) \rangle = -\langle x_i(T) \hat x_i(0) y_j(T') \hat y_j(0)\rangle_S + \langle x_i(T) \hat x_i(0) \rangle_S \langle y_j(T') \hat y_j(0)\rangle_S . 
	\end{align}
	The task now amounts to evaluating the disorder-averaged response functions. We first explore how to do this in the GOE and GUE cases to illustrate the procedure. We then extend the consideration to the general non-Gaussian ensembles in the main text. 
	\section{GOE and GUE cases}
	Although the primary focus of the main text is on real symmetric matrices, of which GOE matrices are an example, it is also helpful to consider the GUE ensemble for the sake of illustrating the diagrammatic formalism. In the GOE case, $\underline{\underline{a}}$ is a symmetric real matrix with Gaussian random entries that have statistics
	\begin{align}
		\langle a_{ij} \rangle = 0, \hspace{1cm} \langle a_{ij}^2\rangle = \langle a_{ij}a_{ji}\rangle = \frac{\sigma^2}{N}. \label{goestats}
	\end{align}
	In the GUE case, one considers complex Hermitian Gaussian random matrices drawn from the ensemble $P(\underline{\underline{a}}) = \mathcal{N}^{-1} \exp\left[-\frac{N}{2 \sigma^2} \sum_{ij} a_{ij}a_{ji} \right]$, where $\mathcal{N}$ is a normalisation constant. Such matrices have statistics
	\begin{align}
		\langle a_{ij} \rangle , \hspace{1cm} \langle a_{ij}^2 \rangle = 0, \hspace{1cm} \langle a_{ij} a_{ji} \rangle = \frac{\sigma^2}{N}.
	\end{align}
	The reason for considering the GUE case here is that it is comparatively simple to evaluate its diagrammatic series, and will provide a helpful comparison for the GOE case. 
	
	\subsection{Disorder-averaged MSRJD actions for GOE and GUE cases}
	To compute the disorder average in Eq.~(\ref{disorderaverageobservable}), we could simply carry out the Gaussian integration for the GOE and GUE cases. However, for the sake of the later calculations and to demonstrate the universality of the results beyond the GOE and GUE ensembles, we proceed slightly differently. 
	
	Letting $f_{ij} = \int dt  \{\hat x_i(t) x_j(t) + \hat y_i(t) y_j(t)\}$, and noting that the pairs $(a_{ij}, a_{ji})$ are independent from one another, the average in Eq.~(\ref{disorderaverageobservable}) can be factorised so that
	\begin{align}
		e^S = e^{S_0} \prod_{i<j}\left\langle\exp\left[ -i(a_{ij} f_{ij} + a_{ji}f_{ji})\right]\right\rangle , \label{factorisation}
	\end{align}
	where we define the `bare' action
	\begin{align}
		S_0 =  i\sum_{i} \int dt \, \hat x_i (\dot x_i  + \omega x_i) + \hat y_i (\dot y_i  + \mu y_i). \label{S0}
	\end{align}
	We can therefore expand the exponential in Eq.~(\ref{factorisation}) and take the disorder average to obtain in the GOE case (invoking the symmetry $a_{ij}=a_{ji}$)
	\begin{align}
		&\left\langle\exp\left[-i a_{ij} (f_{ij} +f_{ji})\right] \right\rangle = 1 -\frac{i}{1!} \left \langle a_{ij} \right\rangle (f_{ij}+f_{ji}) - \frac{1}{2!}   \left\langle a_{ij}^2 \right\rangle (f_{ij}+f_{ji})^2 + \cdots \nonumber \\
		&= 1 - \frac{\sigma^2}{2! N} (f_{ij}+f_{ji})^2  + \cdots \approx \exp\left[- \frac{\sigma^2}{2! N}  (f_{ij}+f_{ji})^2\right], \label{exponexpan}
	\end{align}
	where we have used the fact that $N \gg 1$. This leads us to the GOE action
	\begin{align}
		S_{\mathrm{GOE}} = - \frac{\sigma^2}{2 \times 2!} \frac{1}{N} \sum_{ij} \int dt \int dt' &\left[ \hat x_i(t) x_j(t) + \hat y_i(t) y_j(t) + \hat x_j(t) x_i(t) + \hat y_j(t) y_i(t)  \right]\nonumber \\
		&\times\left[ \hat x_i(t') x_j(t') + \hat y_i(t') y_j(t') + \hat x_j(t') x_i(t') + \hat y_j(t') y_i(t')  \right], \label{GOE}
	\end{align}
	where we include contributions $i = j$ in the above sum, which contribute a negligible $O(1/N)$ term. In the GUE case, we instead obtain
	\begin{align}
	S_{\mathrm{GUE}} = - \frac{\sigma^2}{ 2!}\frac{1}{N} \sum_{ij} \int dt \int dt' &\left[ \hat x_i(t) x_j(t) + \hat y_i(t) y_j(t)   \right]\left[ \hat x_j(t') x_i(t') + \hat y_j(t') y_i(t')  \right] . \label{GUEaction}
	\end{align}	
	One notes that we would have arrived at the same expressions if we were to consider non-Gaussian random matrix ensembles with higher moments that decayed more quickly than $1/N$, which would have resulted in negligible higher-order contributions to the action. We therefore immediately see that there is a universality of the GOE/GUE results. The results for the ensembles in the main text do not possess such sufficiently quickly decaying higher-order statistics, which is why they do not belong to this universality class.
	
	\subsection{Rainbow diagrams, 1-point Green's function, and semi-circle law}\label{section:semicircle}
	To evaluate the response functions that we desire, we now make the following observation, letting $S = S_0 + S_\mathrm{int}$ and considering a general $S_\mathrm{int}$,
	\begin{align}
		N^{-1}\sum_{k} \left\langle R^{(x)}_{kk}(T,0) \right \rangle= -i 	N^{-1}\sum_{k}\sum_{r} \left\langle \frac{S_\mathrm{int}^r}{r!} x_k(T) \hat x_k(0) \right\rangle_0, \label{expsum}
	\end{align}
	where $\langle \cdot \rangle_0$ indicates an average with respect to the bare action, i.e.
	\begin{align}
		\langle O \rangle_0 = \int D[x,\hat x, y, \hat y] \, O \, e^{S_0} . 
	\end{align}
	We evaluate each of the terms in the sum in Eq.~(\ref{expsum}) using Wick's theorem. Wick's theorem is valid for averages with respect to the bare action, which is quadratic in the dynamic variables. More precisely, the average of an even number of the dynamic variables is given by the sum of all possible combinations of the variables averaged in pairs. For example, the average of four dynamic variables with respect to the bare action simplifies as follows
	\begin{align}
		\langle x_k(t)\hat x_k(t') x_i(T) \hat x_{i}(T')\rangle_0 =& \langle x_k(t)\hat x_k(t') \rangle_0 \langle x_i(T) \hat x_{i}(T')\rangle_0 + \langle x_k(t)\hat x_i(T') \rangle_0 \langle x_i(T) \hat x_{k}(t')\rangle_0 \nonumber \\
		& + \langle x_k(t) x_i(T) \rangle_0 \langle \hat x_i(t') \hat x_{k}(T')\rangle_0  \nonumber \\
		=& \langle x_k(t)\hat x_k(t') \rangle_0 \langle x_i(T) \hat x_{i}(T')\rangle_0 + \langle x_k(t)\hat x_i(T') \rangle_0 \langle x_i(T) \hat x_{k}(t')\rangle_0 ,
	\end{align}
	where we note that $\langle x_k(t) x_i(T) \rangle_0 \langle \hat x_i(t') \hat x_{k}(T')\rangle_0 = 0$ [see Eq.~(\ref{hatzero})]. Crucially, we have the following relation for the bare response function
	\begin{align}
		-i\langle x_i(t) \hat x_{j}(T) \rangle_0 = R^{(0)}_{ij} (t,T) = \delta_{ij} e^{- \omega(t-T)} \Theta(t-T) , 
	\end{align}
	where $\Theta(\cdot)$ is the Heaviside function, meaning that we can evaluate averages with respect to the bare action explicitly.
	
	Keeping track of the huge variety of `Wick pairings' in the sum in Eq.~(\ref{expsum}) is a daunting task. A useful strategy is therefore to represent the non-vanishing terms as a series of Feynman diagrams. Aside from the identity in Eq.~(\ref{hatzero}), terms can also vanish due to the time ordering of the dynamic variables $x_i(t)$ and $\hat x_i(t')$ since $R^{(0)}_{ij}(t, t') = 0$ for $t<t'$ due to causality. They can also vanish in the thermodynamic limit $N \to \infty$, since $R^{(0)}_{ij}(t,t') \propto \delta_{ij}$ and therefore some Wick pairings will be subleading in $1/N$ once we carry out the sums over the indices $i,j, \cdots$. These simplifications manifest diagrammatically. In general, only planar diagrams \cite{brezin1978planar, t1993planar} (with no crossing arcs) that consist of one connected piece (no time loops) survive. We refer the reader to the pedagogical Ref. \cite{hertz2016path} for more information on this topic.
	
As simple examples, let us consider the Feynman diagrams that arise from the $r = 1$ term in Eq.~(\ref{expsum}) in the GOE and GUE cases. In the GUE case, only the following diagram is non-zero
	\begin{figure}[H]
		\centering 
		\includegraphics[scale = 0.4]{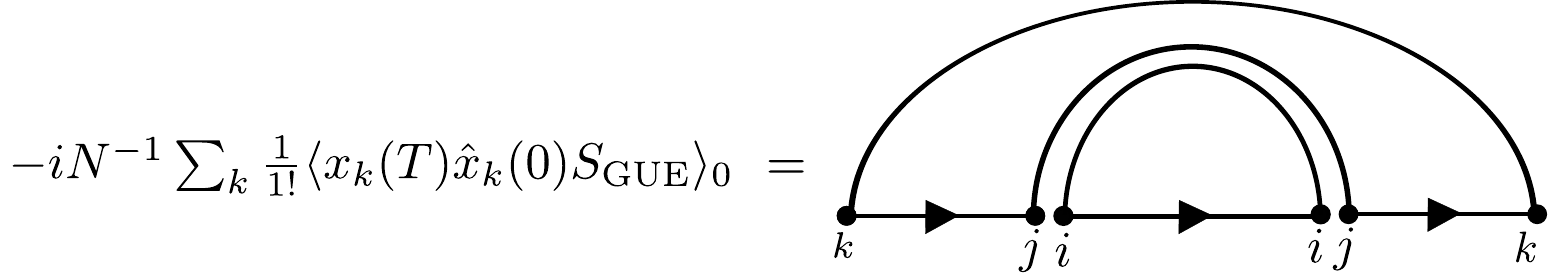}
		\captionsetup{labelformat=empty}
	\end{figure}
The above diagram should be interpreted as follows (see also Ref. \cite{baron2022eigenvalues}):  A dot on the left-hand end of a directed edge represents an $\hat x$-type variable, and a dot on the right-hand end of a directed edge represents an $x$-type variable. Pairs of dots positioned together have the same time coordinate. The $x$ and $\hat x$ variables connected by an arc are constrained to have the same index. Double arcs carry a multiplicative factor of $\sigma^2/N$. Points connected by horizontal edges are Wick-paired together (averaged with respect to the bare action), and thus evaluate to the bare response function. Because $R^{(0)}_{ij}(t,t') = 0$ for $t<t'$, the time coordinate of an $x$-type variable must always be greater than that of an $\hat x$-type variable, hence the directionality of the edges. Finally, all internal times (i.e. not corresponding to the nodes at either end of the diagram) and all indices are summed/integrated over. That is, one should read the above diagram as 
\begin{align}
-iN^{-1} \sum_{k} \frac{1}{1!} \langle x_k(T)	\hat x_k(0) S_\mathrm{GUE}\rangle_0 &= \frac{\sigma^2}{N^2} \sum_{i,j,k} \frac{1}{1!}  (-i)^3 \int dt \int dt' \langle x_k(T) \hat x_j(t) \rangle_0 \langle x_i(t) \hat x_i(t') \rangle_0 \langle x_j(t') \hat x_k(0) \rangle_0 \nonumber \\
&= \frac{\sigma^2}{N^2} \sum_{i,j,k} \int dt \int dt' R_{kj}(T,t) R_{ii}(t,t') R_{jk}(t',0), \nonumber \\
&= \frac{\sigma^2}{N^2} \sum_{ijk} \delta_{kj} \delta_{jk} \delta_{ii} \int^T_0 dt \int_{0}^t dt' e^{-\omega(T-t)} e^{-\omega(t-t')} e^{-\omega t'} \nonumber \\
&=\sigma^2 \int^T_0 dt \int_{0}^t dt' e^{-\omega(T-t)} e^{-\omega(t-t')} e^{-\omega t'} , \label{firstorderexample}
\end{align}
where we note that a combinatorial factor $1/2!$ has cancelled due to the symmetry $(i\leftrightarrow j,t \leftrightarrow t')$. We also note that for the purposes of calculating the 1-point response functions, the $y$-type auxilliary variables in action Eq.~(\ref{GUEaction}) can be ignored, because they give rise only to Feynman diagrams with closed time loops, which vanish due to causality. We note that the integral in Eq.~(\ref{firstorderexample}) is a convolution, and so the Laplace transform is evaluated easily
\begin{align}
	\lim_{\eta \to 0}\mathcal{L}_T\left\{-iN^{-1} \sum_{k} \frac{1}{1!} \langle x_k(T)	\hat x_k(0) S_\mathrm{GUE}\rangle_0 \right\}(\eta) = \frac{\sigma^2}{\omega^3} . 
\end{align}

On the other hand, we have the following diagrams for the GOE
	\begin{figure}[H]
	\centering 
	\includegraphics[scale = 0.4]{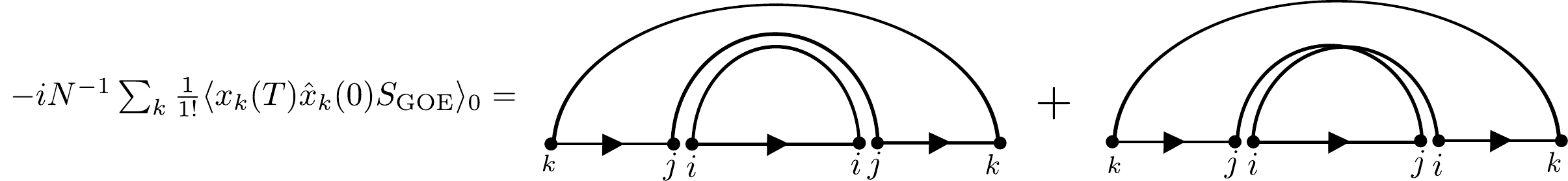}
	\captionsetup{labelformat=empty}
\end{figure}
In general, diagrams are proportional to $N^{E - A/2 -1}$, where $E$ is the number of disconnected (by arcs) sets of directed horizontal edges, and $A$ is the number of double arcs (i.e. excluding the one connecting the end points). This means that the right-hand diagram with the twisted double arc is $O(N^{-1})$, and therefore negligible in the thermodynamic limit. Because the contribution of diagrams with twisted double arcs is negligible, the GOE and GUE have the same diagrammatic series for the 1-point Green's function. 

Let us take the additional example of the set of diagrams that arise from the $r= 2$ term in Eq.~(\ref{expsum}) to further elucidate why we focus only on planar diagrams in the limit $N\to \infty$. The second order term in $S_\mathrm{GUE}$ has the following surviving terms
\begin{align}
	&-\frac{i}{2! } N^{-1} \sum_k\langle x_k(T) \hat x_k(0) S_{\mathrm{GUE}}^2\rangle_0 \nonumber \\
	&= \frac{\sigma^4}{N^3}\int dt_1 dt'_1 dt_2 dt'_2 \sum_{k,i_1, j_1,i_2, j_2}(-i)^6 \Bigg[\nonumber \\
	& \left\langle  x_k(T)   \hat x_{i_1}(t_1) \right\rangle_0  \left\langle x_{j_1}(t_1)    \, \hat x_{j_1}(t'_1) \right\rangle_0   \left\langle x_{i_1}(t_1')    \, \hat x_{i_2}(t_2) \right\rangle_0 \left\langle x_{j_2}(t_2)    \, \hat x_{j_2}(t'_2) \right\rangle_0   \left\langle   x_{i_2} (t'_2) \hat x_k(0) \right\rangle_0  \nonumber \\
	&+ \left\langle  x_k(T)   \hat x_{i_1}(t_1) \right\rangle_0  \left\langle x_{j_1}(t_1)    \,  \hat x_{i_2}(t_2) \right\rangle_0 \left\langle x_{j_2}(t_2)  \,  \hat x_{j_2}(t'_2) \right\rangle_0  \left\langle   x_{i_2} (t'_2)  \, \hat x_{j_1}(t'_1) \right\rangle_0    \left\langle x_{i_1}(t_1') \hat x_k(0) \right\rangle_0 \nonumber \\
	&+\left\langle  x_k(T)   \hat x_{i_1}(t_1) \right\rangle_0  \left\langle x_{j_1}(t_1)    \, \hat x_{i_2}(t_2) \right\rangle_0 \left\langle x_{j_2}(t_2)   \, \hat x_{j_1}(t'_1) \right\rangle_0   \left\langle x_{i_1}(t_1')   \, \hat x_{j_2}(t'_2) \right\rangle_0   \left\langle   x_{i_2} (t'_2) \hat x_k(0) \right\rangle_0
	\Bigg]   , \label{secondordersurviving}
\end{align}
and we have used that there is a symmetry between the times labelled $1$ and $2$ (which has cancelled the factor of $2!$ from Eq.~(\ref{expsum})) and symmetry between dashed and undashed times (which has cancelled a factor of $(2!)^2$ from $S_\mathrm{GUE}^2$). We note that due to this kind of symmetry, the specific labelling of the vertices in the diagrams is irrelevant. The number of ways of ordering the times always cancels the appropriate multiplicative factor, and so the only salient feature of a diagram is its topology \cite{hertz2016path}. 

Only the first two of these Wick pairings survives in the thermodynamic limit. This can be seen simply by observing 
\begin{align}
	&\frac{1}{N}\sum_{k} \lim_{\eta\to 0}\mathcal{L}_T\left\{-\frac{i}{2! }\langle x_k(T) \hat x_k(0) S_{\mathrm{GUE}}^2\rangle_0\right\}(\eta) \nonumber \\
	&= \frac{\sigma^4}{\omega^5}\Bigg[\frac{1}{N^3} \sum_{k, i_1, j_1, i_2, j_2}  \delta_{k,i_1} \delta_{j_1,j_1} \delta_{i_1,i_2} \delta_{j_2,j_2} \delta_{j_2,k}	 \nonumber \\ 
	&+\frac{1}{N^3} \sum_{k, i_1, j_1, i_2, j_2}   \delta_{k,i_1} \delta_{j_1,i_2} \delta_{j_2,j_2} \delta_{i_2,j_1}\delta_{i_1,k}  \nonumber \\
	&+\frac{1}{N^3} \sum_{k, i_1, j_1, i_2, j_2}   \delta_{k,i_1} \delta_{j_1,i_2} \delta_{j_2,j_1} \delta_{i_1,j_2} \delta_{i_2,k} \Bigg]	,
\end{align} 
where we see that the first two products of Kronecker deltas evaluate to $1$, whereas the final set gives $1/N^2$. 

The real advantage of the diagrammatic representation is in identifying those Wick pairings that vanish in the same way that the third pairing in Eq.~(\ref{secondordersurviving}) did. The Wick pairings in Eq.~(\ref{secondordersurviving}) can be represented diagrammatically as 
\begin{figure}[H]
	\centering 
	\includegraphics[scale = 0.45]{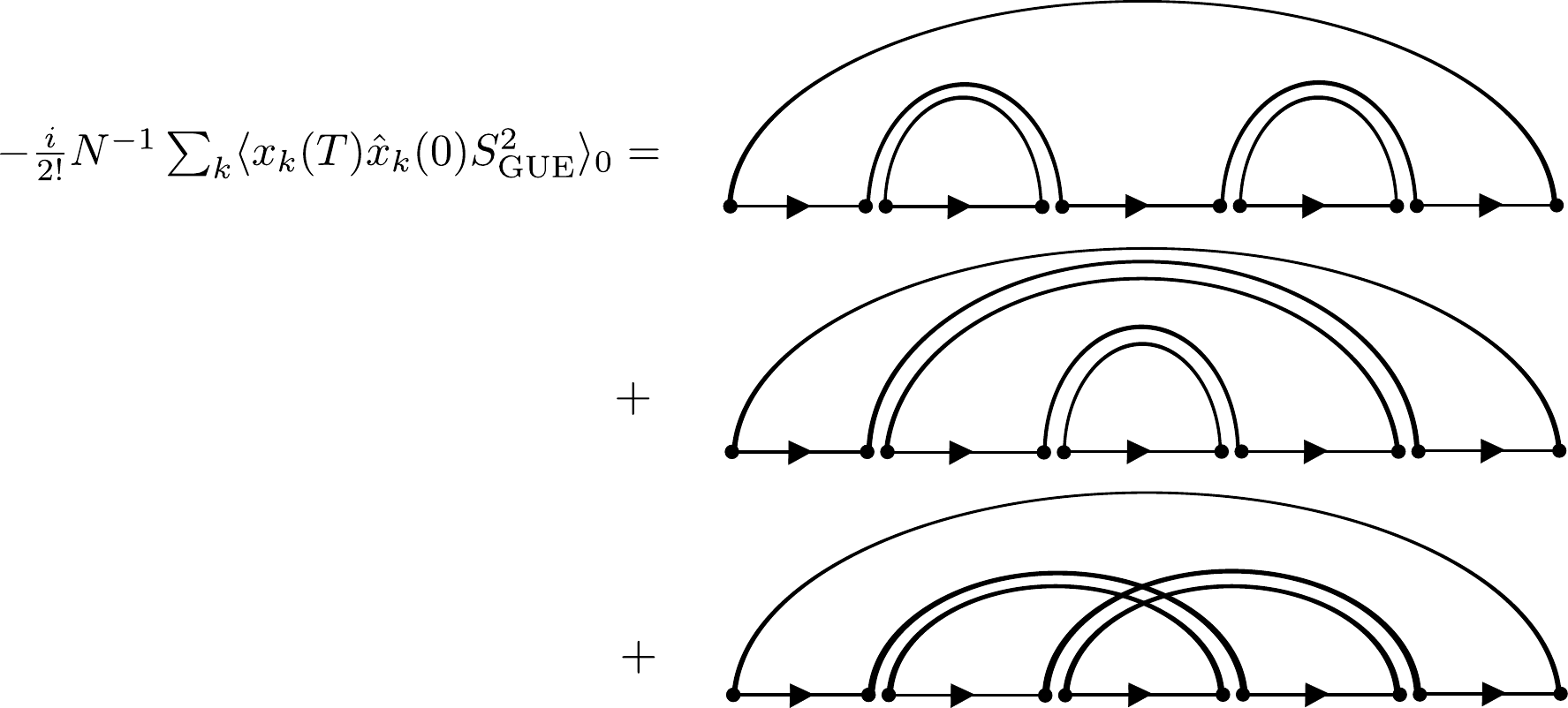}
	\captionsetup{labelformat=empty}
\end{figure}
These first two digrams each have three disconnected sets of directed edges, which corresponds to three factors of $\sum_l \delta_{ll}$. This cancels the factor of $N^{-3}$. In contrast, the third term in Eq.~(\ref{secondordersurviving}) is represented by a diagram whose directed edges are all connected by arcs. This means that one obtains only a single factor of $\sum_{l} \delta_{ll}$ after summing over all other indices. One thus finds that this diagram is an $O(N^{-2})$ contribution. In the limit $N\to \infty$, we therefore have
\begin{align}
	&\frac{1}{N}\sum_{k} \lim_{\eta\to 0}\mathcal{L}_T\left\{-\frac{i}{2! }\langle x_k(T) \hat x_k(0) S_{\mathrm{GUE}}^2\rangle_0\right\}(\eta) = \frac{2\sigma^4}{\omega^5}	,
\end{align} 

We thus see that the `non-planar' diagram gives a contribution that vanishes in the limit $N\to \infty$ and only the planar diagrams survive. We also saw that a factor of $1/(2! (2!)^2)$ cancelled due to time ordering. Indeed, the surviving diagrams are identical once again in the GOE case, with the only difference between the two cases being additional diagrams with twisted arcs, which are subleading in $1/N$.

To summarise, we have so far argued that the following simplifying rules apply generally: 
\begin{enumerate}
	\item The only Wick pairings that we need to consider pair solely hatted and unhatted dynamic variables.
	\item The only non-vanishing Wick pairings for $N\to \infty$ correspond to planar diagrams with non-crossing and non-twisted arcs.
	\item The number of combinations of Wick pairings that are equivalent up to time ordering always exactly cancels a prefactor, allowing us to discard the labelling of the internal nodes in the Feynman diagrams.
\end{enumerate}

One therefore sees that the sum in Eq.~(\ref{expsum}) can be evaluated in the thermodynamic limit by considering the set of all planar rainbow diagrams. As a final example, we find the following non-vanishing diagrams for the third-order term in both the GUE and GOE cases (where $S_\mathrm{int}$ here stands for either $S_\mathrm{GUE}$ or $S_\mathrm{GOE}$)
\begin{figure}[H]
	\centering 
	\includegraphics[scale = 0.45]{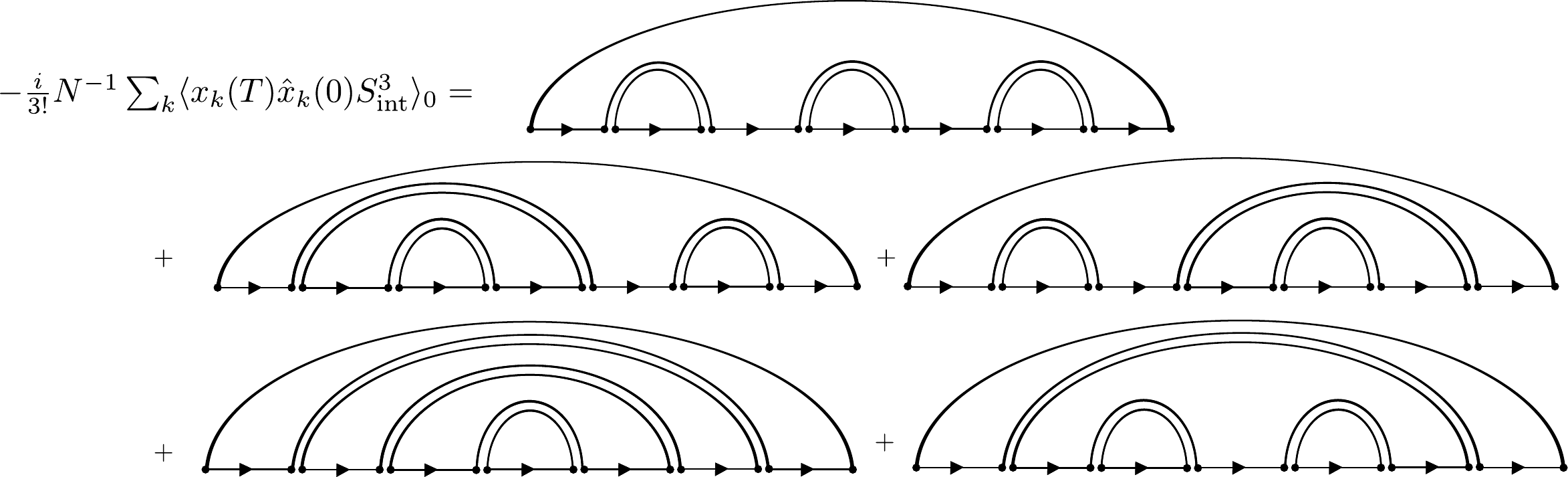}
	\captionsetup{labelformat=empty}
\end{figure}
These diagrams give us 
\begin{align}
	&\frac{1}{N} \sum_k \lim_{\eta\to 0}\mathcal{L}_T \left\{-\frac{i}{3!}\left\langle x_k(T) \hat x_k(0) S_\mathrm{int}^3 \right\rangle_0 \right\}(\eta) = \frac{5\sigma^6}{\omega^7}.
\end{align}
We thus see how the formidable task of evaluating the series in Eq.~(\ref{expsum}) simplifies to summing a series of planar diagrams, each of which can be evaluated in terms of elementary functions. 

We employ one additional diagrammatic convention to simplify the notation when we perform sums over many diagrams. We denote a sum of planar diagrams by an edge with a double arrow, accompanied by a label for identification purposes. For example, let us take the surviving planar diagrams for the second-order term above $-\frac{i}{2!}  \langle x_k(t) \hat x_l S_\mathrm{int}^2\rangle_0 \equiv (O_2)_{kl}$, for which we write
\begin{figure}[H]
	\centering 
	\includegraphics[scale = 0.45]{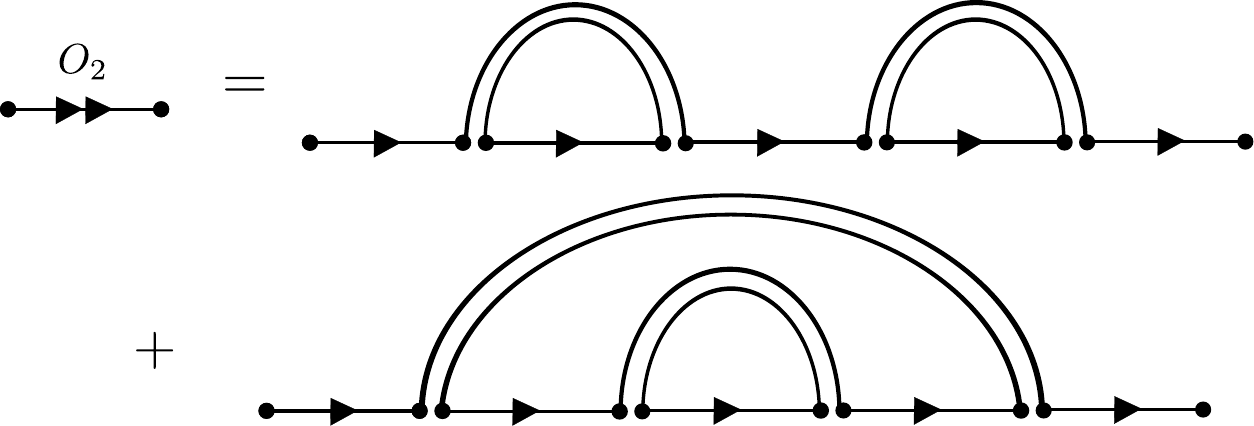}
	\captionsetup{labelformat=empty}
\end{figure}
When we draw an arc over a double-arrowed edge, this is also to be interpreted as a sum of diagrams. Precisely, for the example above we have
\begin{figure}[H]
	\centering 
	\includegraphics[scale = 0.45]{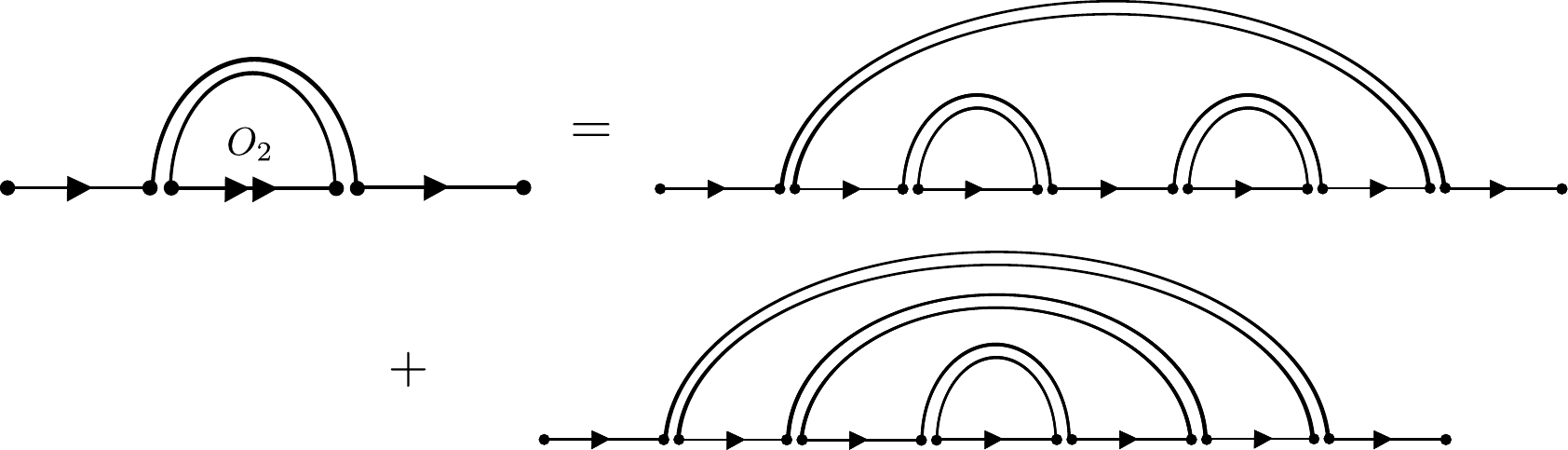}
	\captionsetup{labelformat=empty}
\end{figure}

\begin{figure}[H]
	\centering 
	\includegraphics[scale = 0.25]{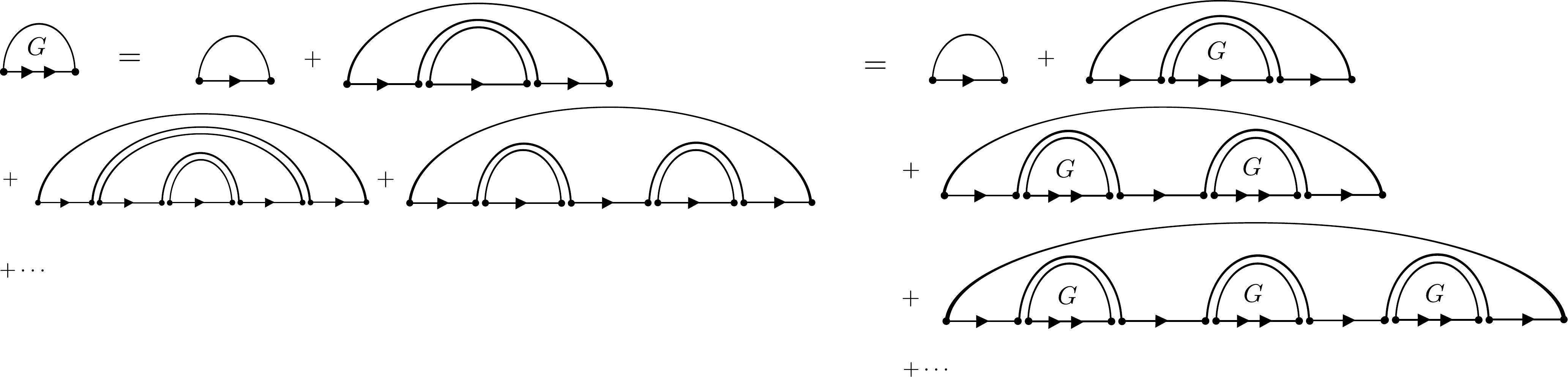}
	\caption{The sum over all possible diagrams. Recognising the self-similarity of the series, this can be rewritten as a geometric series. }\label{fig:rainbowdiagrams}
\end{figure}

We use this convention in Fig. \ref{fig:rainbowdiagrams} below to sum the full series of planar diagrams. We summarise the argument briefly here as to why the two series in Fig. \ref{fig:rainbowdiagrams} are same. 

Let us say that a diagram has $r_e$ `external arcs' if, by following a completely connected path of vertices from the leftmost vertex to the rightmost, we traverse $r_e$ arcs. We can categorise a general planar rainbow diagram by the number of external arcs that it has, since no arcs intersect. The full collection of diagrams with a single external arc, for example, can then be found by taking every planar diagram in the series, placing each of them inside a single arc, and attaching two directed edges to either side. Similar statements apply for diagrams with any number of external arcs. The complete series of planar diagrams can therefore be generated by summing together all of the sets of diagrams with $r_e = 1, 2, 3, \cdots$ external arcs, where under each arc is the sum of all diagrams in the series. In this statement, we have thus identified a self-similarity quality of the series, which allows us to perform the resummation.

Because of this argument, we thus see why the full series of rainbow diagrams can be represented by the simpler series involving $G(\omega)$ (the dressed resolvent) in Fig. \ref{fig:rainbowdiagrams}. This simpler series is recognised to be geometric and is given by
\begin{align}
	&G(\omega) =  \lim_{\eta\to 0}N^{-1} \sum_k\mathcal{L}_T\left[-i\langle x_k(T) \hat x_k(0)  \rangle_S \right](\eta)= \frac{1}{\omega} + \frac{\sigma^2}{\omega^2} G(\omega) + \frac{\sigma^4}{\omega^3} G^2(\omega) + \cdots . \label{selfconsistentseriesexplicit}
\end{align}
This series can be resummed to give
\begin{align}
	G(\omega) = \frac{1}{\omega - \sigma^2G(\omega)}. 
\end{align}
Finally, using Eq.~(\ref{densfromres}), we obtain the Wigner semi-circle law, which is valid for $N\to \infty$ for both the GOE and GUE ensembles,
\begin{align}
	\rho(\omega) = \frac{1}{2 \pi \sigma^2}\sqrt{4 \sigma^2 - \omega^2} .
\end{align}
\subsection{Ladder diagrams, 2-point Green's functions, and eigenvalue correlations}
Let us now turn our attention to the 2-point eigenvalue correlations, which are accessible from the path-integral formalism via the 2-point Green's function in Eq.~(\ref{2pointgc}). 

We use the same trick as in the case of the 1-point functions, and expand the exponential term containing the interaction action, arriving at a similar series to Eq.~(\ref{expsum}). Explicitly, we have
\begin{align}
	N^{-2}\sum_{k,l} \left\langle R^{(x)}_{kk}(T,0) R^{(y)}_{ll}(T',0) \right \rangle = - 	N^{-2}\sum_{k, l}\sum_{r} \left\langle \frac{S_\mathrm{int}^r}{r!} x_k(T) \hat x_k(0)  y_l(T') \hat y_l(0) \right\rangle_0, \label{expsum2point}
\end{align}
where $S_\mathrm{int}$ is given either by $S_\mathrm{GOE}$ or $S_\mathrm{GUE}$ in Eqs.~(\ref{GOE}) or (\ref{GUEaction}) respectively. We then use Wick's theorem to evaluate the infinite series of terms. In this case, since we have both $x$- and $y$-type variables, the diagrammatic representation is more complicated. We depict the $x$-propagator as a directed horizontal edge pointing rightwards, and we depict the $y$-propagator as left-pointing horizontal edge. Some example diagrams are as follows in the case of the GOE

\begin{figure}[H]
	\centering 
	\includegraphics[scale = 0.3]{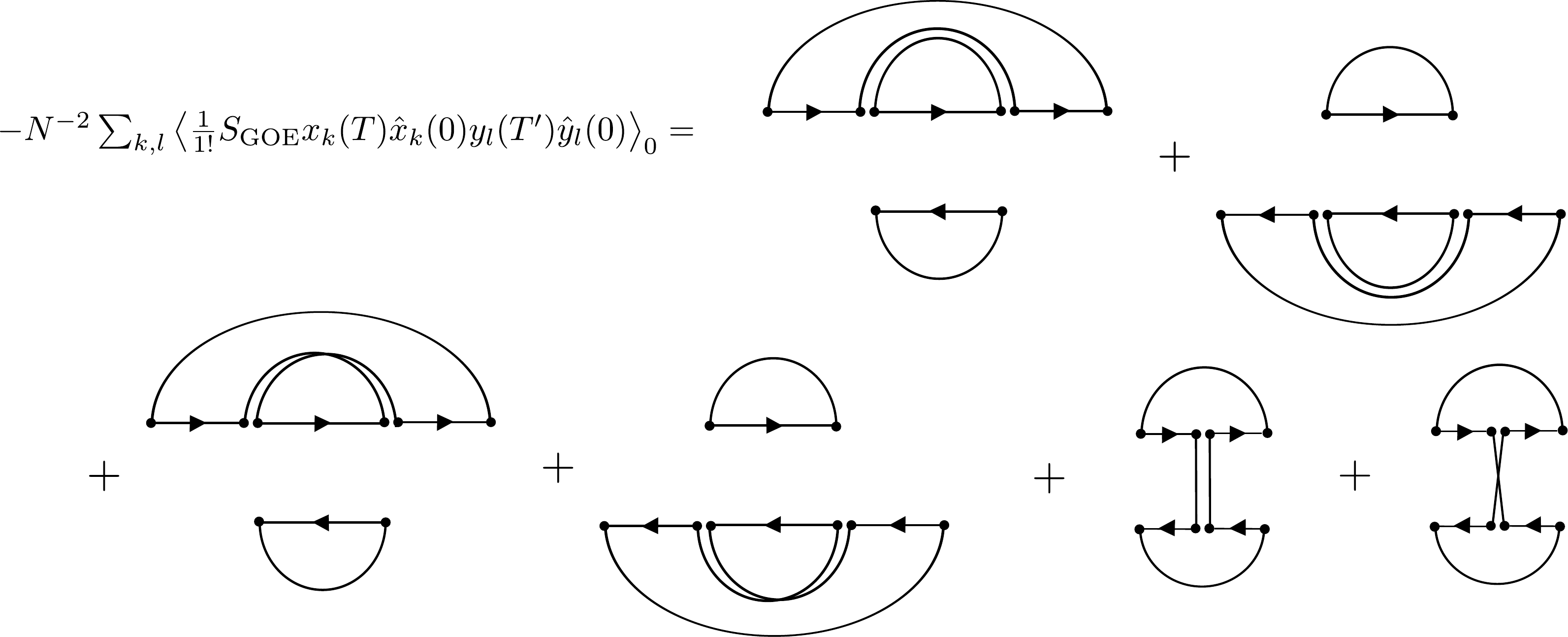}
	\label{fig:ladderdiagramsexample}
\end{figure}

In the case of the GUE, we have similar diagrams, but without twisted arcs

\begin{figure}[H]
	\centering 
	\includegraphics[scale = 0.3]{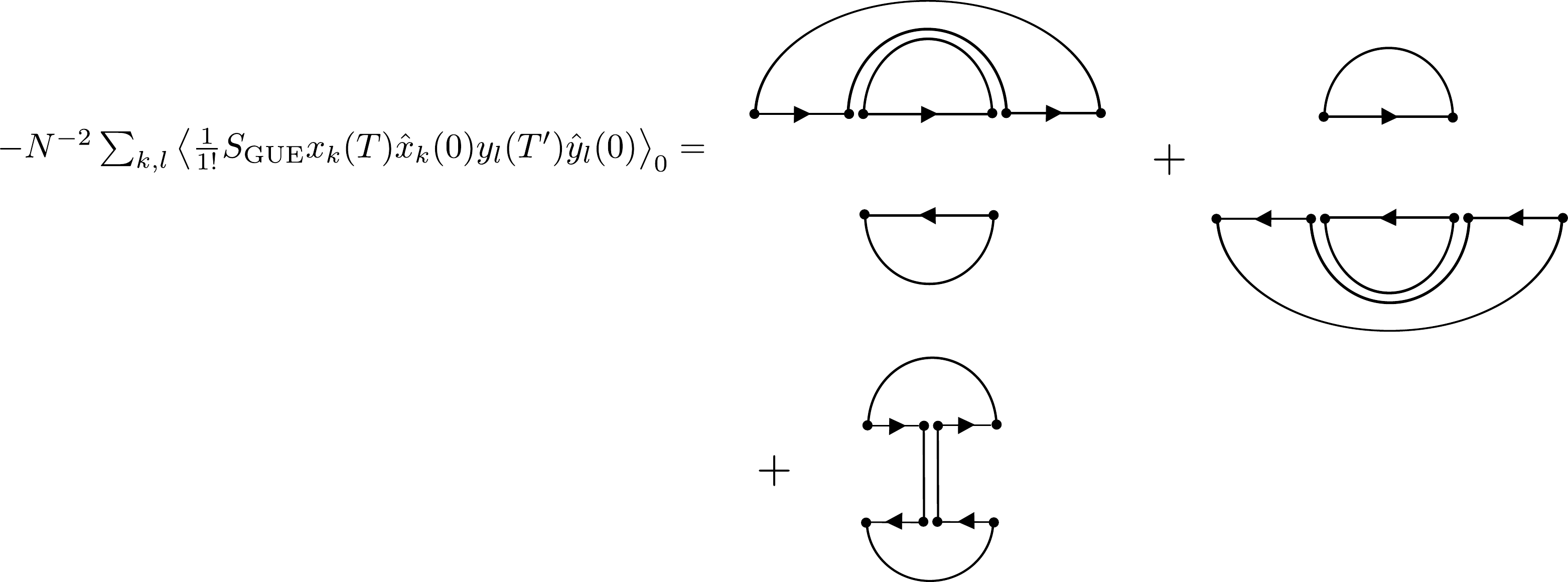}
\label{fig:ladderdiagramsexampleherm}
\end{figure}

In evaluating the \textit{connected} 2-point function in Eq.~(\ref{2pointgc}), we are taking the difference between two diagrammatic series. One finds that the only diagrams that survive are those that are not simply the product of rainbow diagrams. That is, only \textit{connected} diagrams survive, which are known as ladder diagrams. 

\pagebreak

In the GUE case, the full series can be respresented as
	\begin{figure}[H]
		\centering 
		\includegraphics[scale = 0.25]{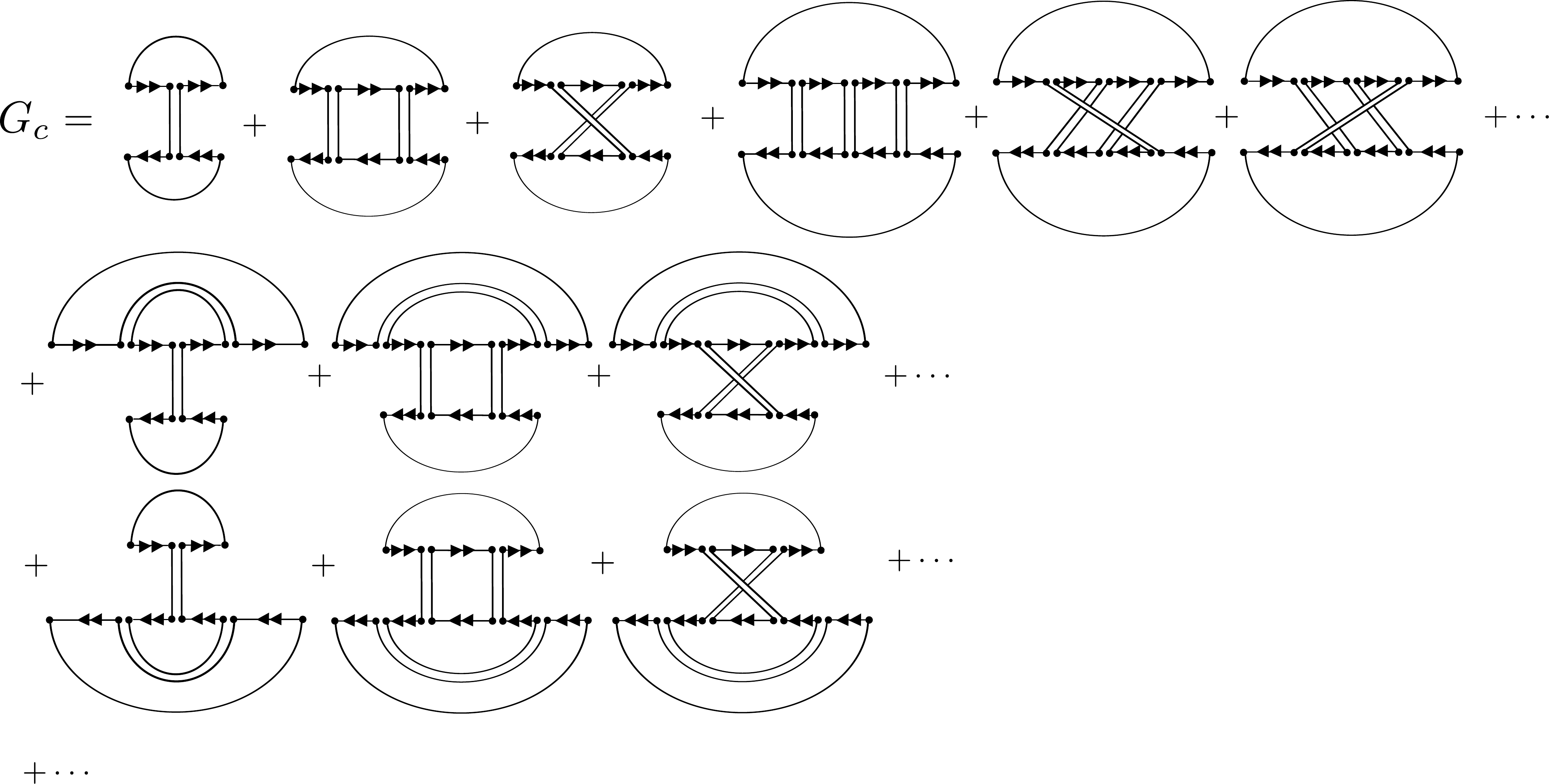}
\label{fig:ladderdiagrams}
	\end{figure}
Here, we have once again used double-arrow notation to represent the 1-point Green's function, foregoing the label $G$ to save space. We have already recognised that in the sum over all possible diagrams, it is always possible to replace any propagator line with the full series of 1-point rainbow diagrams.

We see here that the leading order of diagram is now $\propto 1/N^2$. As per Ref. \cite{brezin1994correlation}, we can resum the diagrammatic series to leading order in $1/N$. This is accomplished by rearranging the series into subseries, as above. The crucial thing to note is that diagrams with $n$ sets of vertical double lines can be cyclically permuted $n$ times. One can also place arbitrarily many arcs over all the double lines on either the top or bottom to produce the other subseries on the lines below. Resumming the full series, one thus obtains
\begin{align}
	G_c(\omega, \mu) = \frac{1}{N^2}\frac{1}{\left[ 1- \sigma^2 G(\omega)G(\mu)\right]^2}\frac{\sigma G^2(\omega)}{1- \sigma^2 G^2(\omega)}\frac{ \sigma G^2(\mu)}{1- \sigma^2 G^2(\mu)},
\end{align}
and we therefore have using Eq.~(\ref{corrfromgc})
\begin{align}
	\rho_c(\omega,\mu) = -\frac{1}{2 N^2 \pi^2}\frac{1}{(\omega-\mu)^2} \frac{4\sigma^2 - \omega\mu}{\sqrt{(4\sigma^2 - \omega^2)(4\sigma^2- \mu^2)}} .
\end{align}
One notes here that when $\vert \omega-\mu \vert \sim N^{-1}$ (i.e. the separation is of the order of the average eigenvalue spacing) this result breaks down. This is because the series can no longer be treated perturbatively in $1/N$, and we would have to sum over all orders in $1/N$ (i.e. more complicated diagrammatic topologies) in order to obtain a valid result. For $\vert \omega-\mu \vert \sim N^{-1}$, universal Wigner-Dyson statistics instead apply, which are better treated using other methods such as  orthogonal polynomials \cite{brezin1993universality} or the supersymmetric approach \cite{verbaarschot1985critique,mirlin1991universality,efetov1983supersymmetry}. 

It is not so simple to resum the series in the GOE case where there are twisted arcs. We can only say that the eigenvalue density correlations will be $O(N^{-2})$ in this case also. It has been found previously by other methods \cite{verbaarschot1984replica} that we also have in the GOE case
\begin{align}
	\rho_c(\omega, \mu)   \sim& \frac{-1}{N^2 (\omega-\mu)^2}, \label{goemacrotwopoint}
\end{align}
for macroscopic eigenvalue separations $\vert \mu - \omega\vert \gg [N \rho([\mu+\omega]/2)]^{-1}$.  However, in both cases, we see that long-range eigenvalue correlations essentially vanish in the thermodynamic limit. In particular, one finds that the contribution of long-range correlations to the eigenvalue compressibility [see Eq.~(\ref{compressibilitydef})] is vanishing. That is, the eigenvalue compressibility is dominated by the short-range eigenvalue repulsion, which is described by universal Wigner-Dyson statistics.

\section{Perturbations about the GOE action}
Having discussed how diagrammatic methods can be employed to derive the 1- and 2-point eigenvalue density statistics in the standard GOE case, we now extend the discussion to the more general ensembles of the main text. We reiterate that the kinds of non-Gaussian statistics that we consider here are
\begin{align}
	\langle a_{ij}^2 \rangle = \frac{\sigma^2}{N} , \hspace{1cm} \langle a_{ij}^2 a_{ik}^2 \rangle - \langle a_{ij}^2 \rangle \langle a_{ik}^2 \rangle  = \frac{\alpha_\mathrm{het} \sigma^4}{N^2} , \hspace{1cm}
	\langle a_{ij}^4 \rangle = \frac{\alpha_4 \sigma^4}{N} , \hspace{1cm} 
	\langle a_{ij}a_{jk}a_{ki} \rangle = \frac{\alpha_\mathrm{cyc} \sigma^3}{N^2}. \label{stats}
\end{align} 
We show here that these statistics give rise to an MSRJD action of the form
\begin{align}
	S \approx S_0 + S_\mathrm{GOE} + \alpha_4 S_\mathrm{4} + \alpha_\mathrm{het} S_\mathrm{het} + \alpha_\mathrm{cyc} S_\mathrm{cyc} + O(\alpha^2), \label{genericaction}
\end{align}
where $S_0$ is as in Eqs.~(\ref{S0}), and we have defined [where here again $f_{ij} =\int dt \{\hat x_i(t) x_j(t) + \hat y_i(t) y_j(t)\} $]
\begin{align}
	S_\mathrm{GOE} &= -\frac{\sigma^2}{2\times2! N} \sum_{ij} (f_{ij} + f_{ji})^2 =  -\frac{\sigma^2}{2\times2! N} \sum_{ij} \int dt_1dt_2 [\hat x_i(t_1) x_j(t_1) + \hat x_j(t_1) x_i(t_1) +\hat y_i(t_1) y_j(t_1) + \hat y_j(t_1) y_i(t_1) ] \nonumber \\
	&\hspace{7cm}\times[\hat x_i(t_2) x_j(t_2) + \hat x_j(t_2) x_i(t_2) +\hat y_i(t_2) y_j(t_2) + \hat y_j(t_2) y_i(t_2) ] , \nonumber \\
	S_\mathrm{4} &= \frac{\sigma^4}{2 \times 4!\, N} \sum_{i,j} (f_{ij} +f_{ji})^4 = \frac{\sigma^4}{2 \times 4!\, N} \sum_{i,j} \int dt_1 \cdots dt_4 [\hat x_i(t_1) x_j(t_1) + \hat x_j(t_1) x_i(t_1) +\hat y_i(t_1) y_j(t_1) + \hat y_j(t_1) y_i(t_1) ] \nonumber \\
	&\hspace{5cm}\times\cdots\times[\hat x_i(t_4) x_j(t_4) + \hat x_j(t_4) x_i(t_4) +\hat y_i(t_4) y_j(t_4) + \hat y_j(t_4) y_i(t_4) ]  , \nonumber \\
	S_\mathrm{het} &= \frac{\sigma^4}{ 2\times(2!)^2 N^2}  \sum_{i,j,k} (f_{ij} +f_{ji})^2(f_{ik} + f_{ki})^2 \nonumber \\
	S_\mathrm{cyc} &= i\frac{\sigma^3}{3! \, N^2} \sum_{i,j,k} (f_{ij} +f_{ji})(f_{jk} + f_{kj})(f_{ki} + f_{ik}), \label{actionterms}
\end{align}
where we have neglected to explicit the time integrals in the last two expressions, which take a similar form first two. We can show that if a generic ensemble has an MSRJD action of this form, then it must possess matrix elements with the statistics in Eq.~(\ref{stats}). One can see this because the MSRJD generating functional is related simply to the moment generating function of the joint distribution of the matrix entries (where here $f_{ij}$ are taken to be parameters)
\begin{align}
	F(\{f_{ij}\}) = \left\langle e^{-i \sum_{ij} f_{ij} a_{ij}}\right\rangle = \exp\left[ S - S_0\right],
\end{align}
so that we have (assuming $\frac{\partial F}{\partial f_{ij}}\Big\vert_{f= 0} = 0$)
\begin{align}
	\langle a_{ij}^2 \rangle = -\frac{\partial^2 F}{\partial f_{ij}^2}\Big\vert_{f= 0}, &\hspace{1cm} \langle a_{ij}^2 a_{ik}^2 \rangle - \langle a_{ij}^2 \rangle \langle a_{ik}^2 \rangle  = \frac{\partial^4 F}{\partial f_{ij}^2 \partial f_{ik}^2}\Big\vert_{f= 0} - \frac{\partial^2 F}{\partial f_{ij}^2}\Big\vert_{f= 0} \frac{\partial^2 F}{ \partial f_{ik}^2}\Big\vert_{f= 0}  , \nonumber \\
	\langle a_{ij}^4 \rangle = \frac{\partial^4 F}{ \partial f_{ij}^4}\Big\vert_{f= 0} , &\hspace{1cm} 
	\langle a_{ij}a_{jk}a_{ki} \rangle = -i\frac{\partial^3 F}{\partial f_{ij} \partial f_{jk}\partial f_{ki}}\Big\vert_{f= 0} . \label{statsfromgf}
\end{align} 
By simple differentiation, we thus see that the action in Eqs.~(\ref{genericaction}) and (\ref{actionterms}) recovers the statistics in Eq.~(\ref{stats}). 

To illustrate the converse (i.e. that the action of the form in Eq.~(\ref{genericaction}) is indeed the correct one, and that there are no other relevant terms to consider that contribute at the leading order in $\alpha$), we evaluate the action in the case of several ensembles. We see that the $\alpha$ parameters that emerge correspond meaningfully to a parameter that encodes a first-order deviation from GOE statistics. These examples will later also be used to verify the general theory. Later, from the generic MSRJD action, we then proceed to calculate the 1- and 2-point Green's functions using the same methodology as the GOE case, arriving eventually at the modified semi-circle law and expressions for eigenvalue density correlations and the compressibility presented in the main text.

\subsection{Example ensembles and corresponding MSRJD actions}

\subsubsection{Matrices with long-tailed distributions and sparse Erd\H{o}s-R\'enyi graphs }
Let us present here examples of ensembles that exhibit $\alpha_4 \neq 0$ and $\alpha_\mathrm{het} = \alpha_\mathrm{cyc} = 0$ (with $\alpha_4$, $\alpha_\mathrm{het}$ and $\alpha_\mathrm{cyc}$ being as defined in Eq.~(1) of the main text). Consider first sparse random matrices $\underline{\underline{a}}$ whose non-zero elements represent a weighted Erd\H{o}s-R\'enyi graph. In other words, a link between two nodes $i$ and $j$ exists with probability $p/N$. If a link between the two nodes exists, we draw $a_{ij}$ from a distribution $\pi(a_{ij})$, and set $a_{ji} = a_{ij}$. All other entries of $\underline{\underline{a}}$ are set to zero. The joint distribution of the matrix elements $a_{ij}$ and $a_{ji}$ is therefore given by 
\begin{align}
	P(a_{ij}, a_{ji}) =\delta (a_{ij}- a_{ji})\left[ \left(1 - \frac{p}{N}\right) \delta(a_{ij}) + \frac{p}{N} \pi(a_{ij})\right] \label{sparsedef}
\end{align}
We see readily that $p$ is the mean number of connections per node on the network (i.e. the average number of non-zero random matrix elements per row/column). We denote the lower-order statistics of the distribution~$\pi(a_{ij})$~by
\begin{align}
	\left\langle a_{ij} \right\rangle_\pi = \frac{\mu}{p}, \hspace{1cm}
	\left\langle (a_{ij}- \mu/p)^2\right\rangle_\pi = \frac{\sigma^2}{p}, \hspace{1cm}
	\langle (a_{ij}-\mu/p)^4 \rangle_\pi = \frac{\Gamma_4 \sigma^4}{p^2},  \label{loworderstats}
\end{align}
where $\langle \cdot \rangle_\pi$ indicates an average over the distribution $\pi$ (to be contrasted with $\langle\cdots\rangle$, which denotes an average over realizations of the network \textit{and} the weights of links). Scaling the variance of the interaction coefficients with $p$ as in Eq.~(\ref{loworderstats}) permits one to take the dense limit $p\to N$ in a sensible fashion. It also allows one to perform a perturbative expansion in powers of $1/p$ transparently. The scaling can easily be undone by substituting $\sigma^2 \to p \sigma^2$ and $\mu \to p \mu$. We also assume that higher order statistics scale with higher powers of $1/p$ such that $\langle a_{ij}^6 \rangle_\pi \sim 1/p^3$, and so on. 

For simplicity, we assume that the random variables $a_{ij}$ are centred such that $\mu = 0$. Overall, the statistics of $\{a_{ij}\}$ (not conditioned on there being a link) are thus
\begin{align}
	\left\langle a_{ij} \right\rangle = 0, \hspace{1cm}
	\left\langle a_{ij}^2\right\rangle = \frac{\sigma^2}{N}, \hspace{1cm}
	\langle a_{ij}^4 \rangle = \frac{\Gamma_4 \sigma^4}{p N}. \label{loworderstatser}
\end{align}
We thus see that $\alpha _4 = \Gamma_4/p$ in this case, which is small for $p \gg 1$ (i.e. graphs that are not overly sparse). An expansion for small $\alpha$ is thus a $1/p$ expansion in this case \cite{baron2023pathintegralapproachsparse, rodgers1988density}.

One notes that one can also produce similar non-Gaussian interaction statistics by drawing all elements of the random matrix from a distribution such as a truncated Cauchy-Lorentz distribution
\begin{align}
	P(a_{ij}) = \frac{\epsilon}{\epsilon^2 + (a_{ij} -\mu/N)^2} \frac{1}{\arctan\left( \frac{w}{\epsilon}\right)} \Theta\left( w + \frac{\mu}{N} - \vert a_{ij} \vert\right) , \label{cauchydistribution}
\end{align}
where $\Theta(\cdot)$ is the Heaviside function, and we choose $\epsilon = \frac{\sigma \pi \sqrt{p}}{2 N}$ and $w = \frac{\sigma}{\sqrt{p}}$. The choice of scaling for these parameters with $p$ and $N$ gives rise to statistics of the form in Eq.~(\ref{loworderstats}) for large $N$
\begin{align}
	\overline{ a_{ij} } = \frac{\mu}{N}, \hspace{1cm}
	\overline{ (a_{ij}- \mu/N)^2} = \frac{\sigma^2}{N}, \hspace{1cm}
	\overline{ (a_{ij}-\mu/N)^4 } = \frac{ \gamma_1 \sigma^4}{ p N}, \hspace{1cm}
	\overline{ (a_{ij} - \mu/N)^6} =  \frac{\gamma_2\sigma^6}{ p^2 N}, \cdots \label{cauchystats}
\end{align}
where here $\gamma_r = 1/(2r +1)$, and thus $\alpha_4 = \gamma_1/p$, which is again small for $p \gg1$.

In both of these cases, we can construct the MSRJD action as follows. Beginning once again with the definition of the action in Eq.~(\ref{disorderaverageobservable}), and since the elements $a_{ij}$ are independent up to the symmetry $a_{ij} = a_{ji}$ as they were for the GOE, we once again use the factorisation Eq.~(\ref{factorisation}). However, this time we find [c.f. Eq.~(\ref{exponexpan})], again using the shorthand $f_{ij} = \int dt  (\hat x_i x_j + \hat y_i y_j)$,
\begin{align}
	&\left\langle\exp\left[-i a_{ij} (f_{ij} +f_{ji})\right] \right\rangle = 1 -\frac{i}{1!} \left \langle a_{ij} \right\rangle (f_{ij}+f_{ji}) - \frac{1}{2!}   \left\langle a_{ij}^2 \right\rangle (f_{ij}+f_{ji})^2  + \frac{1}{4!}   \left\langle a_{ij}^4 \right\rangle (f_{ij}+f_{ji})^4+ \cdots \nonumber \\
	&= 1 - \frac{\sigma^2}{2! N} (f_{ij}+f_{ji})^2  + \frac{\alpha_4\sigma^4}{4! N} (f_{ij}+f_{ji})^4 + \cdots \approx \exp\left[- \frac{\sigma^2}{2! N}  (f_{ij}+f_{ji})^2 + \frac{\alpha_4\sigma^4}{4! N} (f_{ij}+f_{ji})^4 \right], \label{exponexpansparse}
\end{align}
where now we ignore terms in the exponent that are subleading in $1/N$ and are of the order $O(\alpha^2)$. We therefore find for the MSRJD action in this case
\begin{align}
	S \approx S_0 + S_\mathrm{GOE} + \alpha_4 S_4 
\end{align}
so that the action is indeed of the form in Eq.~(\ref{genericaction}), with $\alpha_\mathrm{het} = \alpha_\mathrm{cyc} = 0$.

\subsubsection{Statistically heterogeneous matrices and complex networks with degree heterogeneity}\label{subsection:heteroexample}
Let us now examine a case for which $\alpha_\mathrm{het} \neq 0$, while $\alpha_4 = \alpha_\mathrm{cyc}=0$. We suppose that the statistics of the matrix elements themselves vary throughout the matrix such that
\begin{align}
	\langle a_{ij} \rangle &= 0, \nonumber \\
	\langle a_{ij}^2 \rangle & = \frac{k_i k_j\sigma^2}{p^2N} , \label{hetstats}
\end{align}
with all higher moments scaling more quickly with $1/N$. In Ref. \cite{poley2024eigenvalue}, an expansion was performed by assuming that the heterogeneity of these statistics was small. That is, 
\begin{align}
	s^2 = \alpha_\mathrm{het} = \frac{1}{N}\sum_{i} \frac{(k_i - p)^2}{p^2} , \label{ssqdef}
\end{align}  
was taken to be a small parameter, where the $k_i$ themselves are random variables with mean $p$. A simple example would be to draw each $k_i$ from a uniform distribution with variance $s^2 p^2$ and mean $p$, and then draw $\{a_{ij}\}$ as independent Gaussian random variables.

Another example of such an ensemble would be the weighted adjacency matrix of a network constructed according to the configuration model or the Chung-Lu model \cite{newman2018networks,chung2002connected}. For the Chung-Lu model, we imagine that a network has an expected degree sequence $\{k_i\}$, where $\{k_i\}$ are themselves random variables drawn from some degree distribution $P_\mathrm{deg}(k)$. Let the mean degree be $p$ and the total number of nodes be $N$, as above. In the Chung-Lu model, a link between nodes $i$ and $j$ exists with probability $\frac{k_i k_j}{pN}$ (which is assumed to be less than 1 for all $i,j$). If a link between nodes $i$ and $j$ exists, $a_{ij}$ is drawn from a distribution with statistics given by Eq.~(\ref{loworderstats}) (with $\mu = 0$) and we set $a_{ji} = a_{ij}$. If a link does not exist, we set $a_{ij} = a_{ji} = 0$. We thus have $\langle a_{ij}^2\rangle = \frac{k_i k_j \sigma^2}{p^2N}$ as required. This also holds true for the configuration model in the regime $1 \ll p \ll N$ \cite{baron2022networks} and is known as the annealed network approximation. The quantity $s^2$ above is then the degree heterogeneity.

Importantly, in both of these cases, $a_{ij} = a_{ji}$ are all still independent random variables for a given set of $k_i$. Thus, to perform the average in Eq.~(\ref{disorderaverageobservable}), we can perform two averages sequentially: one over $a_{ij}$ for a given $k_i$ (which we denote $\langle\cdot\rangle_a$), and one over the $k_i$ themselves (which we denote $\langle\cdot\rangle_k$). We thus have 
\begin{align}
	&\left\langle\exp\left[-i a_{ij} (f_{ij} +f_{ji})\right] \right\rangle_a = 1 -\frac{i}{1!} \left \langle a_{ij} \right\rangle_a (f_{ij}+f_{ji}) - \frac{1}{2!}   \left\langle a_{ij}^2 \right\rangle_a (f_{ij}+f_{ji})^2 + \cdots \nonumber \\
	&= 1 - \frac{k_i k_j\sigma^2}{2! p^2 N} (f_{ij}+f_{ji})^2  + \cdots \approx \exp\left[- \frac{k_i k_j\sigma^2}{2! p^2 N}  (f_{ij}+f_{ji})^2\right]. \label{exponexpanhet}
\end{align}
We then perform the average over the $\{k_i\}$ as follows (for $\delta k_i = k_i - p$)
\begin{align}
	& \left\langle \exp\left[- \sum_{i,j}\frac{k_i k_j\sigma^2}{2 \times 2! p^2 N}  (f_{ij}+f_{ji})^2\right] \right\rangle_k=\left\langle \exp\left[- \sum_{i,j}\frac{\sigma^2}{2 \times 2! N}  (f_{ij}+f_{ji})^2 \left( 1 + 2 \frac{\delta k_i}{p} +\frac{\delta k_i \delta k_j}{p^2}\right)\right] \right\rangle_k \nonumber \\
	 &=  \exp\left[- \sum_{i,j}\frac{\sigma^2}{2 \times 2! N}  (f_{ij}+f_{ji})^2\right] \prod_i\left[1 + \frac{\sigma^4}{(2!)^3 N^2} \sum_{j,k} \left\langle \left(\frac{\delta k_i}{p} \right)^2 \right\rangle_k (f_{ij}+f_{ji})^2 (f_{ik}+f_{ki})^2 + O(\alpha_\mathrm{het}^2) \right] \nonumber \\
	 &\approx \exp\left[- \sum_{i,j}\frac{\sigma^2}{2 \times 2! N}  (f_{ij}+f_{ji})^2 +\frac{\alpha_\mathrm{het} \sigma^4}{(2!)^3 N^2} \sum_{i,j,k}  (f_{ij}+f_{ji})^2 (f_{ik}+f_{ki})^2  \right] . 
\end{align}
We thus obtain an action of the form $S = S_0 + S_\mathrm{GOE} + \alpha_\mathrm{het} S_\mathrm{het}$, which is of the form Eq.~(\ref{genericaction}) with $\alpha_4 = \alpha_\mathrm{cyc} = 0$.
\subsubsection{Cyclic correlations and sparse graphs with loops }\label{subsection:loopsexample}
Let us finally consider a case where $\alpha_\mathrm{cyc} \neq 0$ (as well as $\alpha_4$ and $\alpha_\mathrm{het} \neq 0$), namely the sparse ER graph with additional length-3 loops. More precisely, we suppose that we can decompose $a_{ij}$ as follows
\begin{align}
	a_{ij} = \frac{\sigma}{\sqrt{p+q}}\left[\theta_{(i,j)} +\sum_k \theta_{(i,j,k)} \right] z_{ij} , 
\end{align}
where $\theta_{(i,j)} = 1$ if an edge exists between nodes $i$ and $j$ (which occurs with probability $p/N$) and $\theta_{(i,j,k)} = 1$ if there exists a triangular loop between nodes $i$, $j$ and $k$ (which occurs with probability $q/N^2$). We note that it only occurs with vanishing probability that any one edge is found on two triangular loops or both a normal edge and a loop, but if it is, its weight is simply doubled. The weights $z_{ij} = z_{ji}$ are drawn independently according to the following prescription. Each of four possibilities is equally likely: $(z_{ij}, z_{jk}, z_{ki}) = (1,1,1)$, or $(z_{ij}, z_{jk}, z_{ki}) = (1,-1,-1)$, or $(z_{ij}, z_{jk}, z_{ki}) = (-1,1,-1)$, or $(z_{ij}, z_{jk}, z_{ki}) = (-1,-1,1)$. We also set $(z_{ji}, z_{kj}, z_{ik}) = (z_{ij}, z_{jk}, z_{ki})$. Overall, one has the following statistics (with $\langle a_{ij} \rangle = 0$)
\begin{align}
	\langle a_{ij}^2 \rangle = \frac{\sigma^2}{N}, \hspace{1cm}
	\langle a_{ij}a_{jk}a_{ki} \rangle = \frac{q\sigma^3}{(p + q)^{3/2}N^2}, \hspace{1cm}
	\langle a_{ij}^4 \rangle = \frac{p\sigma^4}{(p+q)^2N}, \hspace{1cm} \langle a_{ij}^2 a_{ik}^2\rangle - \langle a_{ij}^2 \rangle \langle a_{ik}^2\rangle = \frac{q \sigma^4}{N^2 (p+q)^2}. \label{cycstats}
\end{align}
We note that we could simply have set $z_{ji} =  z_{kj} = z_{ik} = z_{ij} = z_{jk} = z_{ki} = 1$, but the above construction yields $\langle z_{ij} \rangle = 0$, which avoids producing a single outlier eigenvalue. So we see that the loops give rise to cyclic correlations between the matrix elements as well as a non-trivial fourth moment, and statistical heterogeneity. In particular, we have here that $\alpha_\mathrm{cyc} = q/(p+q)^{3/2}$, $\alpha_4 = p/(p+q)^2$ and $\alpha_\mathrm{het} = q/(p+q)^2$, which are again all small for large average degree. We note that there also exist algorithms to produce cyclic correlations in dense networks \cite{aceituno2019universal} (i.e. with $\alpha_\mathrm{het} = \alpha_4 = 0$).

We once again must perform the average in Eq.~(\ref{disorderaverageobservable}) over several sets of random variables. We denote the average over the $\theta$ variables by $\langle\cdot \rangle_\theta$, and represent the average of the $z$ variables by $\langle \cdot \rangle_z$. The average in Eq.~(\ref{disorderaverageobservable}) factorises now as follows [c.f. Eq.~(\ref{factorisation})]
\begin{align}
	e^S = e^{S_0} \Bigg\langle &\prod_{(i,j)}\left\langle\exp\left[ -i \frac{\sigma}{\sqrt{p+q}}z_{ij} \theta_{(i,j)}( f_{ij} + f_{ji})\right]\right\rangle_\theta \nonumber \\
	&\times \prod_{(i,j,k)}\left\langle\exp\left[-i \theta_{(i,j,k)} \frac{\sigma}{\sqrt{p+q}}\left[ z_{ij}( f_{ij} + f_{ji}) + z_{jk}( f_{jk} + f_{kj}) + z_{ki}( f_{ki} + f_{ik}) \right]\right]\right\rangle_\theta \Bigg\rangle_z, 
\end{align}
where $\prod_{(i,j)}$ and $\prod_{(i,j,k)}$ denote the product over set of all \textit{combinations} of the indices. Carrying out the averages over the $\theta$ variables, and expanding in $1/\sqrt{p+q}$ as far as $1/(p+q)^2$, one thus has 
\begin{align}
	e^S \approx e^{S_0} \Bigg\langle &\exp\Bigg[ -i\frac{\sigma p}{2 N \sqrt{p+q}} \sum_{i,j} z_{ij} (f_{ij} + f_{ji})- \frac{\sigma^2 p}{2\times 2! N (p+q)} \sum_{i,j} z_{ij}^2 (f_{ij} + f_{ji})^2\nonumber \\
	&+i\frac{\sigma^3 p}{2\times 3! N (p+q)^{3/2}} \sum_{i,j} z_{ij}^3 (f_{ij} + f_{ji})^3 + \frac{\sigma^4 p}{2\times 4! N (p+q)^2} \sum_{i,j} z_{ij}^4 (f_{ij} + f_{ji})^4 \Bigg] \nonumber \\
	&\times \exp\Bigg[ -i\frac{\sigma q}{2 N \sqrt{p+q}} \sum_{i,j} z_{ij} (f_{ij} + f_{ji})- \frac{\sigma^2 q}{2\times 2! N (p+q)} \sum_{i,j} z_{ij}^2 (f_{ij} + f_{ji})^2 \nonumber \\
	&+i\frac{\sigma^3 q}{2\times 3! N (p+q)^{3/2}} \sum_{i,j} z_{ij}^3 (f_{ij} + f_{ji})^3+ \frac{\sigma^4 q}{2\times 4! N (p+q)^2} \sum_{i,j} z_{ij}^4 (f_{ij} + f_{ji})^4 \nonumber \\
	&+ \frac{\sigma^4 q}{(2!)^3 N^2 (p+q)^2} \sum_{i,j,k} z_{ij}^2 z_{ik}^2 (f_{ij} + f_{ji})^2 (f_{ik} + f_{ki})^2 \nonumber \\
	&+ \frac{i\sigma^3 q}{3! N^2 (p+q)^{3/2}} \sum_{i,j,k} z_{ij}z_{jk} z_{ki} (f_{ij} + f_{ji}) (f_{jk} + f_{kj}) (f_{ki} + f_{ik}) + \cdots \Bigg] \Bigg\rangle_z,
\end{align}
where we have omitted some terms involving products like $z_{ij}^2 z_{ki}$ and $z_{ij}^3 z_{ki}$ for the sake of brevity (these terms eventually vanish). Finally, taking the average over the $z$ variables, we obtain
\begin{align}
	S \approx S_0 + S_\mathrm{GOE} + \alpha_4 S_\mathrm{4} + \alpha_\mathrm{het} S_\mathrm{het} + \alpha_\mathrm{cyc} S_\mathrm{cyc} , 
\end{align}
with the $\alpha$ variables taking the values given below Eq.~(\ref{cycstats}).

\subsection{Random matrix ensembles used for figures in main text and Wigner surmise}\label{subsection:ensembles}

In Fig. 2 of the main text, we study 4 different random matrix ensembles. The GOE results are produced from Gaussian random matrices with the statistics as given in Eq.~(\ref{goestats}). The `ER graph' results are produced using matrix elements drawn from the distribution in Eq.~(\ref{sparsedef}) with $p = 30$ and $N = 4000$. This gives rise to a value of $\alpha_\mathrm{4} = 1/p$. The `dense hetero.' ensemble results are produced using matrices constructed according to the Chung-Lu model discussed in Section \ref{subsection:heteroexample}. We use a uniform distribution to draw the values of $k_i$, such that $p - \sqrt{3s^2} p<k_i<p + \sqrt{3s^2}p$. This gives rise to a value of $\alpha_\mathrm{het} = s^2$. We use the values $p = 0.02 N$, $s^2 = 0.03$ and $N = 4000$. Finally, the `ER + loops' ensemble results are produced using the ensemble discussed in Section \ref{subsection:loopsexample}, using $p = 30$, $q = 30$ and $N = 2000$.\\

For the Wigner surmise in the inset of Fig. 2 of the main text, we define the `unfolded' eigenvalue spacings as $\delta_\nu = N \rho[(\lambda_{\nu+1} + \lambda_\nu)/2] (\lambda_{\nu+1}-\lambda_{\nu})$. The GOE has eigenvalue spacings distributed according to the following Wigner surmise \cite{weidenmuller2009random}
\begin{align}
	P(\delta) = \frac{\pi\delta}{2} e^{- \pi \delta^2/4}, \label{surmise}
\end{align}
and we verify that the rest of the ensembles that we consider here do not deviate from this behaviour. The surmise is an excellent approximation for the true GOE level-spacing distribution, which cannot be written down in closed form \cite{gaudin1961loi}.\\

In Fig. 3, we plot the distribution of eigenvector components for matrices of size $N = 10^4$ averaged over 1000 realisations. In the main panel, $a_{ij}$ are each drawn from a separate Gaussian distribution, with a variance that changes depending on the location in the matrix. The variance is given by Eq.~(\ref{hetstats}), where $k_i$ are drawn from a uniform distribution with variance $s^2 p^2 = 0.3 p^2$, and $p = 0.2 N$. In the inset, we use a fully-populated random matrix with i.i.d. elements drawn from a truncated Cauchy distribution defined in Eq.~(\ref{cauchydistribution}), and we use $p = 1.2$. Similar figures to this are given below in Fig. \ref{fig:powerlawtails} for other cases where $\alpha_4 \neq 0$ and $\alpha_\mathrm{het} \neq 0$.

\section{1-point Green's functions and modified Wigner semi-circle}
In Section \ref{section:semicircle}, we produced a diagrammatic series for the 1-point Green's function in the case where the action was $S = S_0 + S_\mathrm{GOE}$, and thereby derived the Wigner semi-circle law. We now wish to compute a similar diagrammatic series, which also takes into account the additional contributions to the action in Eq.~(\ref{genericaction}). 

We begin by writing the series in Eq.~(\ref{expsum}) in a different form for the case of $S_\mathrm{int} = S_\mathrm{GOE} +  \alpha_4 S_\mathrm{4} + \alpha_\mathrm{het} S_\mathrm{het} + \alpha_\mathrm{cyc} S_\mathrm{cyc}$ 
\begin{align}
	N^{-1}\sum_{k} \left\langle R^{(x)}_{kk}(T,0) \right \rangle= -i 	N^{-1}\sum_{k}\sum_{r, r_1, r_2, r_3} \left\langle \frac{(S_\mathrm{GOE})^r (\alpha_4 S_\mathrm{4})^{r_1} (\alpha_\mathrm{het}S_\mathrm{het})^{r_2} (\alpha _\mathrm{cyc}S_\mathrm{cyc})^{r_3}}{r!\, r_1! \,r_2!\, r_3!} x_k(T) \hat x_k(0) \right\rangle_0. \label{expsummod}
\end{align}
We wish only to take into account the first order perturbation in $\alpha$, and therefore, we ignore terms in the above sum for which $r_1 + r_2 + r_3 >1$. The new action terms give rise to new topologies of diagram, with each action contribution having a distinct structure. To illustrate this, let us consider the diagrams for which $r= 0$ and $r_1 + r_2 + r_3 = 1$. The $S_4$ contribution gives rise to a `ribbon' of concatenated arcs \cite{baron2023pathintegralapproachsparse} 
\begin{figure}[H]
	\centering 
	\includegraphics[scale = 0.25]{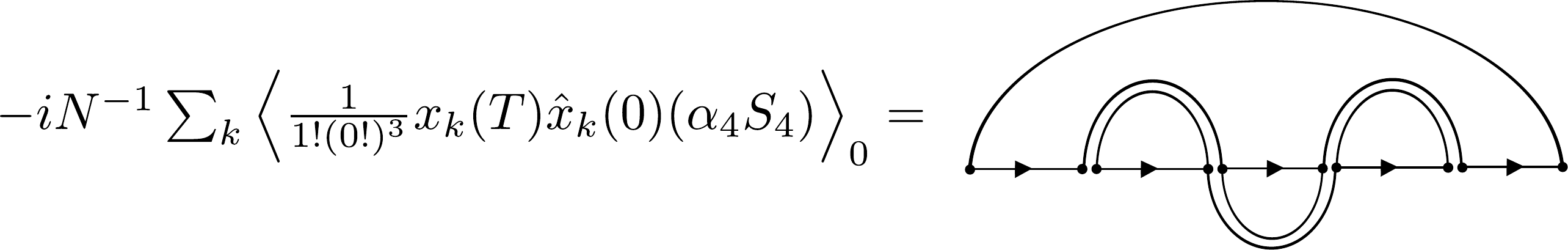}
	\label{fig:examples4}
\end{figure}
We note that $S_4$ contains a factor of $1/N$. We see that the number of disconnected pieces in the above diagram gives a factor of $N^2$, cancelling the factors of $1/N$ from $S_4$ and the prefactor of the expression $N^{-1}\sum_{k} \left\langle R^{(x)}_{kk}(T,0) \right \rangle$. This planar diagram thus survives in the thermodynamic limit. We note that other possible diagrams contributing to $-i N^{-1}\sum_{k} \left\langle \frac{ 1 }{1! (0!)^3} x_k(T) \hat x_k(0) (\alpha_4 S_\mathrm{4}) \right\rangle_0$ have crossing arcs, and thus vanish for $N \to \infty$. As was the case for the rainbow diagrams in Section \ref{section:semicircle}, the combinatorial factor $1/(2\times4!)$ in $S_4$ [see Eq.~(\ref{actionterms})] is cancelled by the number of ways that there are to order the times $t_1, t_2, t_3, t_4$ and the symmetry $i \to j$. The above diagram thus evaluates to
\begin{align}
	\lim_{\eta \to 0}\mathcal{L}_T\left\{-i N^{-1}\sum_{k} \left\langle \frac{ 1 }{1! (0!)^3} x_k(T) \hat x_k(0) (\alpha_4 S_\mathrm{4}) \right\rangle_0 \right\}(\eta)= \frac{\alpha_4\sigma^4}{\omega^5} . 
\end{align}
One instead finds the following diagrams for the $S_\mathrm{het}$ contribution
\begin{figure}[H]
	\centering 
	\includegraphics[scale = 0.2]{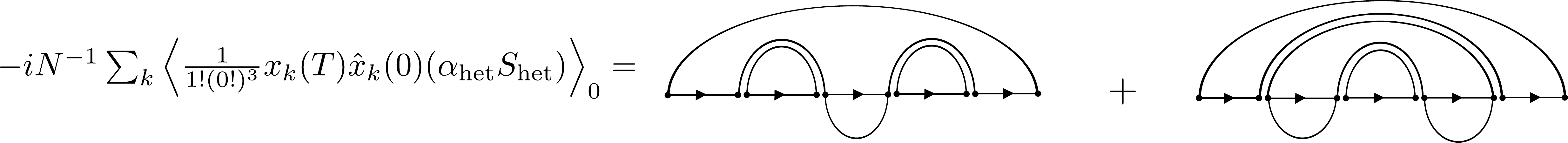}
	\label{fig:ribbondiagramsnetwork}
\end{figure}
where we now see that there is more than one way to arrange the dynamic variables into Wick pairs that gives rise to non-vanishing diagrams. Indeed, the action contribution $S_\mathrm{het}$ contains a factor of $N^{-2}$, and there is an additional factor of $N^{-1}$ in $-i N^{-1}\sum_{k} \left\langle \frac{ 1 }{1! (0!)^3} x_k(T) \hat x_k(0) (\alpha_\mathrm{het} S_\mathrm{het}) \right\rangle_0$. The three disconnected pieces in the above diagrams produce a factor of $N^3$, which cancels the factor of $N^{-3}$, and the diagrams survive in the limit $N \to \infty$. The diagrams thus evaluate to
\begin{align}
	\lim_{\eta \to 0}\mathcal{L}_T\left\{-i N^{-1}\sum_{k} \left\langle \frac{ 1 }{1! (0!)^3} x_k(T) \hat x_k(0) (\alpha_\mathrm{het} S_\mathrm{het}) \right\rangle_0 \right\}(\eta)= \frac{2\alpha_\mathrm{het}\sigma^4}{\omega^5} . 
\end{align}
Finally, one finds the following non-vanishing diagram for the $S_\mathrm{cyc}$ contribution 
\begin{figure}[H]
	\centering 
	\includegraphics[scale = 0.2]{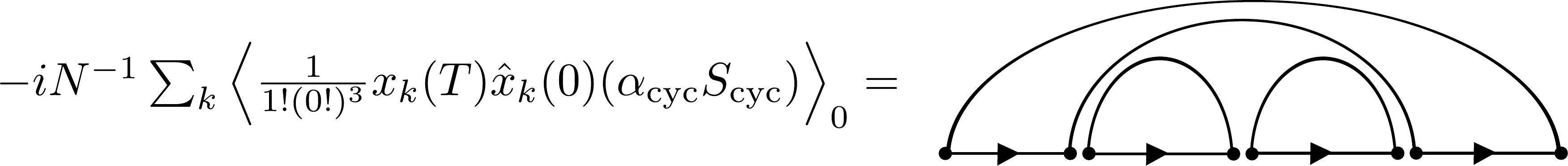}
	\label{fig:cyclicdiagram}
\end{figure}
The action contribution $S_\mathrm{cyc}$ contains a factor of $N^{-2}$, and there is an additional factor of $N^{-1}$ in $-i N^{-1}\sum_{k} \left\langle \frac{ 1 }{1! (0!)^3} x_k(T) \hat x_k(0) (\alpha_\mathrm{cyc} S_\mathrm{cyc}) \right\rangle_0$. The three disconnected pieces in the above diagrams produce a factor of $N^3$ cancels the factor of $N^{-3}$. The diagram thus evaluates to (noting the smaller power of $\omega$)
\begin{align}
	\lim_{\eta \to 0}\mathcal{L}_T\left\{-i N^{-1}\sum_{k} \left\langle \frac{ 1 }{1! (0!)^3} x_k(T) \hat x_k(0) (\alpha_\mathrm{cyc} S_\mathrm{cyc}) \right\rangle_0 \right\}(\eta)= \frac{\alpha_\mathrm{cyc}\sigma^4}{\omega^4} . 
\end{align}
We must now understand how to resum the full series of diagrams with arbitrary $r$ and $r_1 + r_2 + r_3 \leq 1$ [see Eq.~(\ref{expsummod})]. To zeroth order in $\alpha$, one recovers the series of rainbow diagrams of the GOE in Fig. \ref{fig:rainbowdiagrams}. However, we now have additional diagrams such as the following, which are first order in $\alpha_4$
\begin{figure}[H]
	\centering 
	\includegraphics[scale = 0.13]{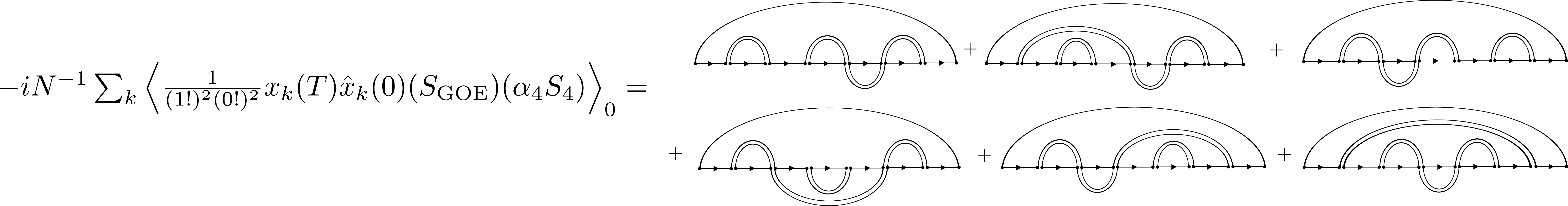}
	\label{fig:exampleribbon}
\end{figure}
The full series of diagrams that are first order in $\alpha_4$ is produced by recognising that the full set of GOE rainbow diagrams can be inserted on every directed edge of the ribbon diagram to produce a diagram that contributes to the series. One can also insert an arbitrary number of double arcs over the ribbon and still maintain the order in $N$. The full series of diagrams of first order in $\alpha_4$ can thus be represented (where we now represent the resolvent for $\alpha = 0$ by $G_0$) 
\begin{figure}[H]
	\centering 
	\includegraphics[scale = 0.3]{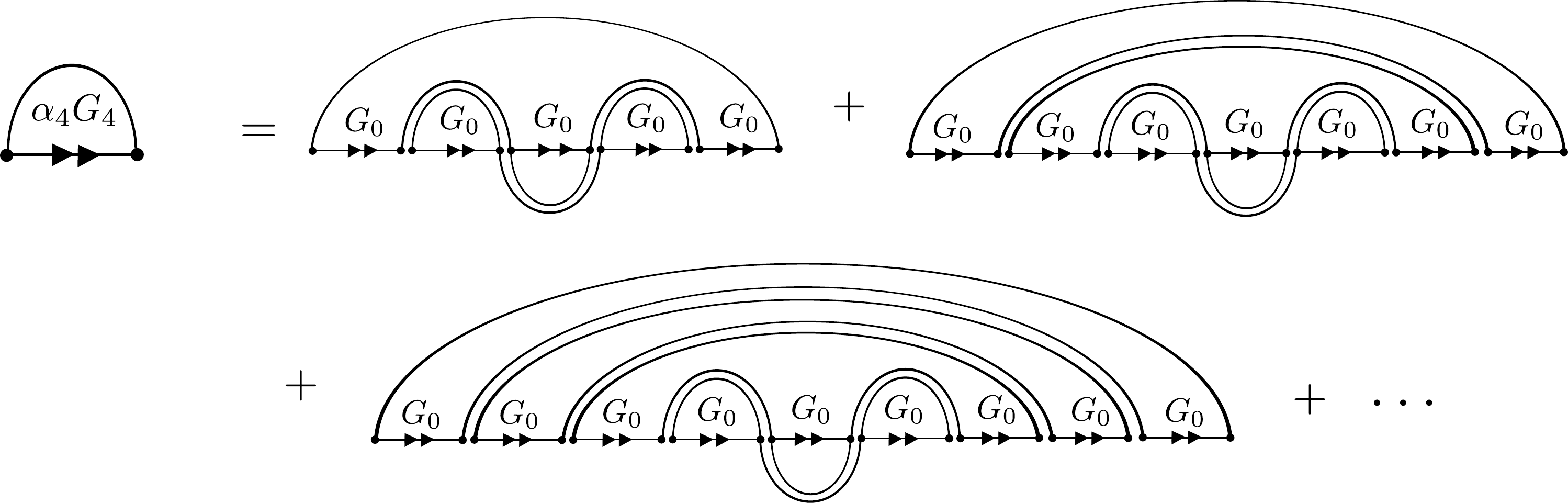}
	\label{fig:ribboncorrection}
\end{figure}
This series thus evaluates to
\begin{align}
	\alpha_4 G_4 = \lim_{\eta \to 0}\mathcal{L}_T\left\{-iN^{-1}\sum_{k}\sum_{r} \left\langle \frac{(S_\mathrm{GOE})^r (\alpha_4 S_\mathrm{4}) }{r!\, 1! \, (0!)^2} x_k(T) \hat x_k(0) \right\rangle_0\right\}(\eta) = \frac{\alpha_4\sigma^4 G_0^5}{1- \sigma^2 G_0^2}. 
\end{align}
We obtain similar series for the other perturbations to the GOE, which similar sum to give
\begin{align}
	\alpha_\mathrm{het} G_\mathrm{het} &= \lim_{\eta \to 0}\mathcal{L}_T\left\{-iN^{-1}\sum_{k}\sum_{r} \left\langle \frac{(S_\mathrm{GOE})^r (\alpha_\mathrm{het} S_\mathrm{het}) }{r!\, 1! \, (0!)^2} x_k(T) \hat x_k(0) \right\rangle_0\right\}(\eta) = \frac{2\alpha_\mathrm{het}\sigma^4 G_0^5}{1- \sigma^2 G_0^2}, \nonumber \\
	\alpha_\mathrm{cyc} G_\mathrm{cyc} &= \lim_{\eta \to 0}\mathcal{L}_T\left\{-iN^{-1}\sum_{k}\sum_{r} \left\langle \frac{(S_\mathrm{GOE})^r (\alpha_\mathrm{cyc} S_\mathrm{cyc}) }{r!\, 1! \, (0!)^2} x_k(T) \hat x_k(0) \right\rangle_0\right\}(\eta) = \frac{\alpha_\mathrm{cyc}\sigma^3 G_0^4}{1- \sigma^2 G_0^2}.
\end{align}
The full diagrammatic series is thus resummed to give
\begin{align}
	G(\omega) = G_0 + \alpha_4 G_4 + \alpha_\mathrm{het} G_\mathrm{het} + \alpha_\mathrm{cyc} G_\mathrm{cyc} = \frac{1}{\omega - \sigma^2 G_0}+ \frac{1}{1- \sigma^2 G_0^2} \left[ \alpha_4\sigma^4 G_0^5 + 2\alpha_\mathrm{het}\sigma^4 G_0^5 +\alpha_\mathrm{cyc}\sigma^4 G_0^4 \right] + O(\alpha^2).
\end{align}
We wish to write this in a self-consistent form. Replacing $G_0 \to G$ in the correction term above only serves to alter the term at next order in $\alpha$, so we write
\begin{align}
	G(\omega)  = \frac{1}{\omega - \sigma^2 G_0}+\sigma^2 G_0^2\left[G(\omega) - \frac{1}{\omega - \sigma^2 G_0} \right] +  \left[ \alpha_4\sigma^4 G^5 + 2\alpha_\mathrm{het}\sigma^4 G^5 +\alpha_\mathrm{cyc}\sigma^3 G^4 \right]+ O(\alpha^2).
\end{align}
Realising that $ \frac{1}{\omega - \sigma^2 G} = \frac{1}{\omega - \sigma^2 G_0} + \sigma^2 \frac{1}{(\omega -\sigma^2 G_0)^2} \left[G(\omega) - \frac{1}{\omega - \sigma^2 G_0} \right] + O(\alpha^2)$, and that $G_0 = \frac{1}{\omega - \sigma^2 G_0}$, we thus arrive at the self-consistent expression that we seek
\begin{align}
G(\omega)  = \frac{1}{\omega - \sigma^2 G} +  \left[ \alpha_4\sigma^4 G^5 + 2\alpha_\mathrm{het}\sigma^4 G^5 +\alpha_\mathrm{cyc}\sigma^3 G^4 \right]+ O(\alpha^2). \label{gselfcons}
\end{align}
We now solve this self-consistent equation along the lines of Ref. \cite{kim1985density} (see also \cite{baron2023pathintegralapproachsparse}). We are careful to correctly preserve the square-root singularity, and thus find the correction not only to the bulk density, but also the spectral edge ($\lambda_\pm$ below).

One notes that $\sigma^2G^2 = \omega G - 1 + O(\alpha)$, so we can rearrange Eq.~(\ref{gselfcons}) to yield the following quadratic 
\begin{align}
	1 - \omega G+ \sigma^2 G^2 + (\alpha_4  + 2\alpha_\mathrm{het})  (\omega G - 1)^2 + \alpha_\mathrm{cyc} \sigma G(\omega G - 1) + O(\alpha^2) = 0.
\end{align}
Solving this now for $G$, we obtain
\begin{align}
	G(\omega) &= \frac{\sigma \alpha_\mathrm{cyc} + [1 + 2(\alpha_4+ 2 \alpha_\mathrm{het})]\omega + \sqrt{(\lambda_- - \omega)(\omega - \lambda_+)}}{2 [\sigma^2 + \sigma \alpha_\mathrm{cyc} \omega + (\alpha_4+ 2 \alpha_\mathrm{het}) \omega^2 ]} + O(\alpha^2), \nonumber \\
	\lambda_\pm &= \pm 2 \sigma [ 1 + \frac{1}{2}( \alpha_4 + 2 \alpha_\mathrm{het} \pm \alpha_\mathrm{cyc}) ]+O(\alpha^2).\label{gpluscorrection}
\end{align}
Upon using Eq.~(\ref{densfromres}), we arrive at 
\begin{align}
	&\rho(\omega) = \frac{-2}{\pi \lambda_+ \lambda_-}\sqrt{(\lambda_+ - \omega)(\omega-\lambda_-)}\bigg[ 1 - 2\alpha_\mathrm{cyc} \frac{\omega}{\sqrt{\left\vert\lambda_+\lambda_-\right\vert}}+(\alpha_4 + 2 \alpha_\mathrm{het})\left( 1 -  \frac{4\omega^2}{\left\vert\lambda_+\lambda_-\right\vert}\right) + O(\alpha^2) \bigg] ,\label{modsemigeneral}
\end{align}
for $\lambda_-<\omega <\lambda_+$. We note that Eq.~(\ref{modsemigeneral}) reduces to the Wigner semi-circle law when $\alpha \to 0$. Eq.~(\ref{modsemigeneral}) is checked against numerics below in Fig. \ref{fig:deltarho}.
\begin{figure}[H]
	\centering 
	\includegraphics[scale = 0.48]{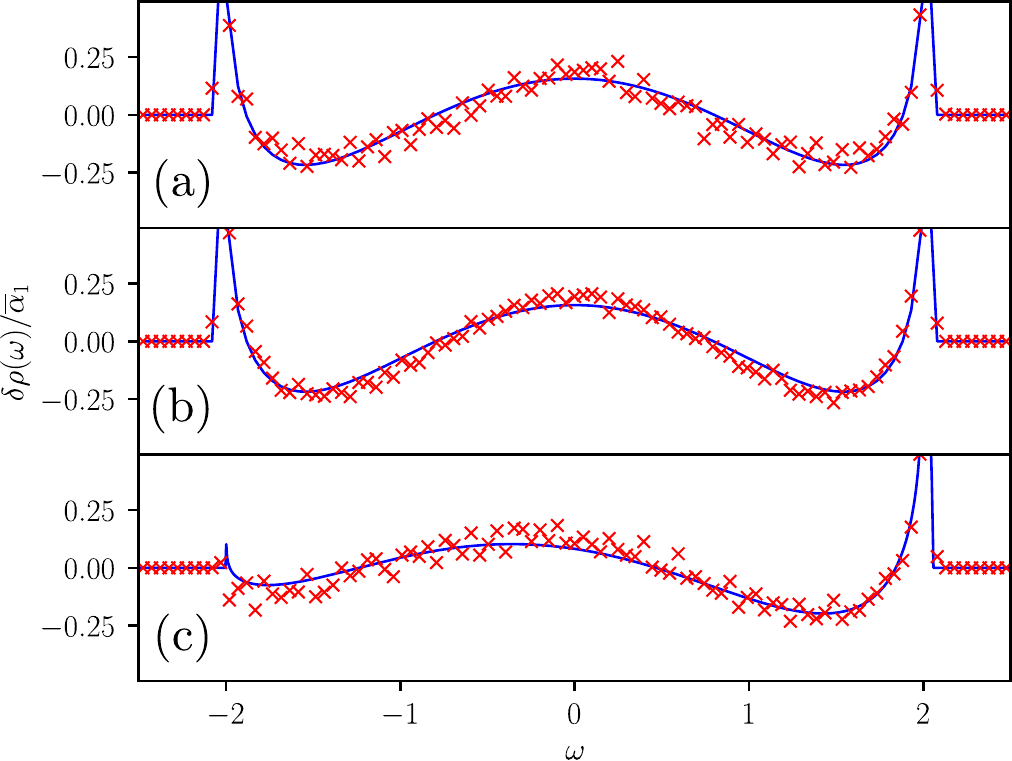}
	\captionsetup{justification=raggedright,singlelinecheck=false, font=small}
	\caption{Deviation from the Wigner semi-circle law in the cases described in Section \ref{subsection:ensembles}. Panel (a): Sparse ER graph with $p = 15$, (b) heterogeneous network with uniform degree distribution and $S = 0.03$ (c) Sparse ER graph with additional loops, with $p = 30$, $q = \sqrt{p}$. The results of numerical diagonalisation are compared with the prediction in Eq.~(\ref{modsemigeneral}). }\label{fig:deltarho}
\end{figure}
\section{2-point Green's functions}

\subsection{$G_c$ and eigenvalue fluctuations $\rho_c$}
Let us now turn to the two-point Green's functions. Using a similar philosophy as for the 1-point functions, we truncate the full series of diagrams at first order in $\alpha$. Precisely, we evaluate the following series diagrammatically
\begin{align}
	&\frac{1}{N^2}\sum_{k,l} \left\langle R^{(x)}_{kk}(T,0) R^{(y)}_{ll}(T',0) \right \rangle - \frac{1}{N^2}\sum_{k,l} \left\langle R^{(x)}_{kk}(T,0) \right\rangle \left\langle R^{(y)}_{ll}(T',0) \right \rangle    \nonumber \\
	&= \frac{-1}{N^2}\sum_{k, l}\sum_{r, r_1, r_2, r_3} \left\langle \frac{(S_\mathrm{GOE})^r (\alpha_4 S_\mathrm{4})^{r_1} (\alpha_\mathrm{het}S_\mathrm{het})^{r_2} (\alpha _\mathrm{cyc}S_\mathrm{cyc})^{r_3}}{r!\, r_1! \,r_2!\, r_3!} x_k(T) \hat x_k(0)  y_l(T') \hat y_l(0) \right\rangle_0 \nonumber \\
	&+\frac{1}{N^2}\sum_{k, l}\sum_{\substack{r, r_1, r_2, r_3 \\ r', r_1', r_2', r_3'}} \left\langle \frac{(S_\mathrm{GOE})^r (\alpha_4 S_\mathrm{4})^{r_1} (\alpha_\mathrm{het}S_\mathrm{het})^{r_2} (\alpha _\mathrm{cyc}S_\mathrm{cyc})^{r_3}}{r!\, r_1! \,r_2!\, r_3!} x_k(T) \hat x_k(0)\right\rangle \nonumber \\
	&\hspace{2cm}\times\left\langle \frac{(S_\mathrm{GOE})^{r'} (\alpha_4 S_\mathrm{4})^{r_1'} (\alpha_\mathrm{het}S_\mathrm{het})^{r_2'} (\alpha _\mathrm{cyc}S_\mathrm{cyc})^{r_3'}}{r'!\, r_1'! \,r_2'!\, r_3'!} y_l(T') \hat y_l(0) \right\rangle_0, \label{expsum2pointmod}
\end{align}
where again we restrict the terms in the sum to $r_1 + r_2 + r_3 \leq 1$. Let us examine some of the contributions that are first-order in $\alpha$. We find that the contribution that is first order in $\alpha_4$ has the following diagrams 
\begin{figure}[H]
	\centering 
	\includegraphics[scale = 0.25]{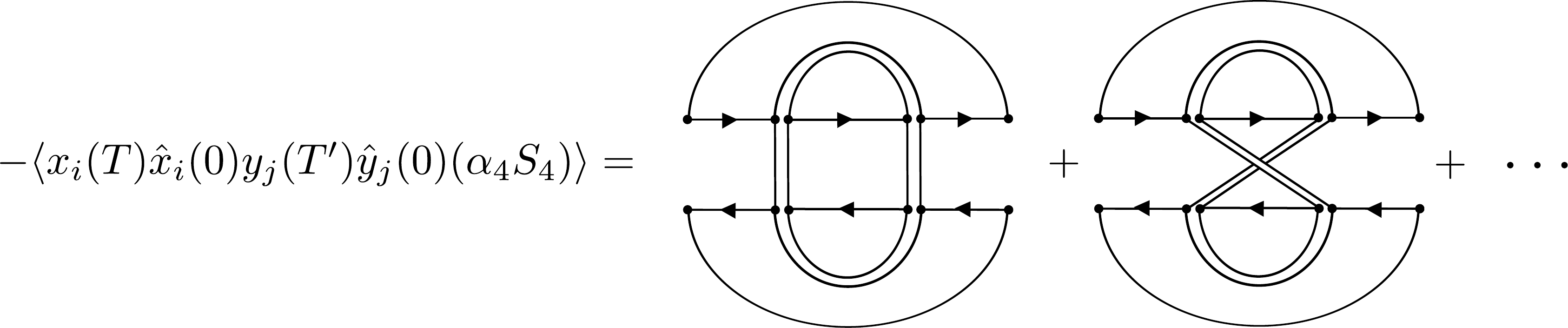}
	\label{fig:ladderdiagramssparseexample}
\end{figure}
where we have excluded unconnected diagrams. One notes the following important point here. The connected diagrams (which are the only ones to contribute to the connected 2-point Green's function in which we are interested), scale differently with $N$ than those of the GOE. One notes that $S_4 \propto 1/N$, and we have an additional factor of $N^{-2}$ in front of $\frac{1}{N^2}\sum_{k,l} \left\langle R^{(x)}_{kk}(T,0) R^{(y)}_{ll}(T',0) \right \rangle$, so in total we have a factor of $1/N^3$. However, the connected diagrams above each have two disconnected pieces, so the order of these diagrams is therefore $N^2/N^3 = 1/N$. That is, they are a factor of $N$ larger than the corresponding diagrams for the GOE. 

Ignoring the diagrammatic contributions from the GOE action (and other diagrams that are $O(1/N^2)$), we can therefore evaluate to leading order in $1/N$ the subseries of connected diagrams that is $\propto\alpha_4$. The full series of relevant diagrams can be represented as follows
\begin{figure}[H]
	\centering 
	\includegraphics[scale = 0.15]{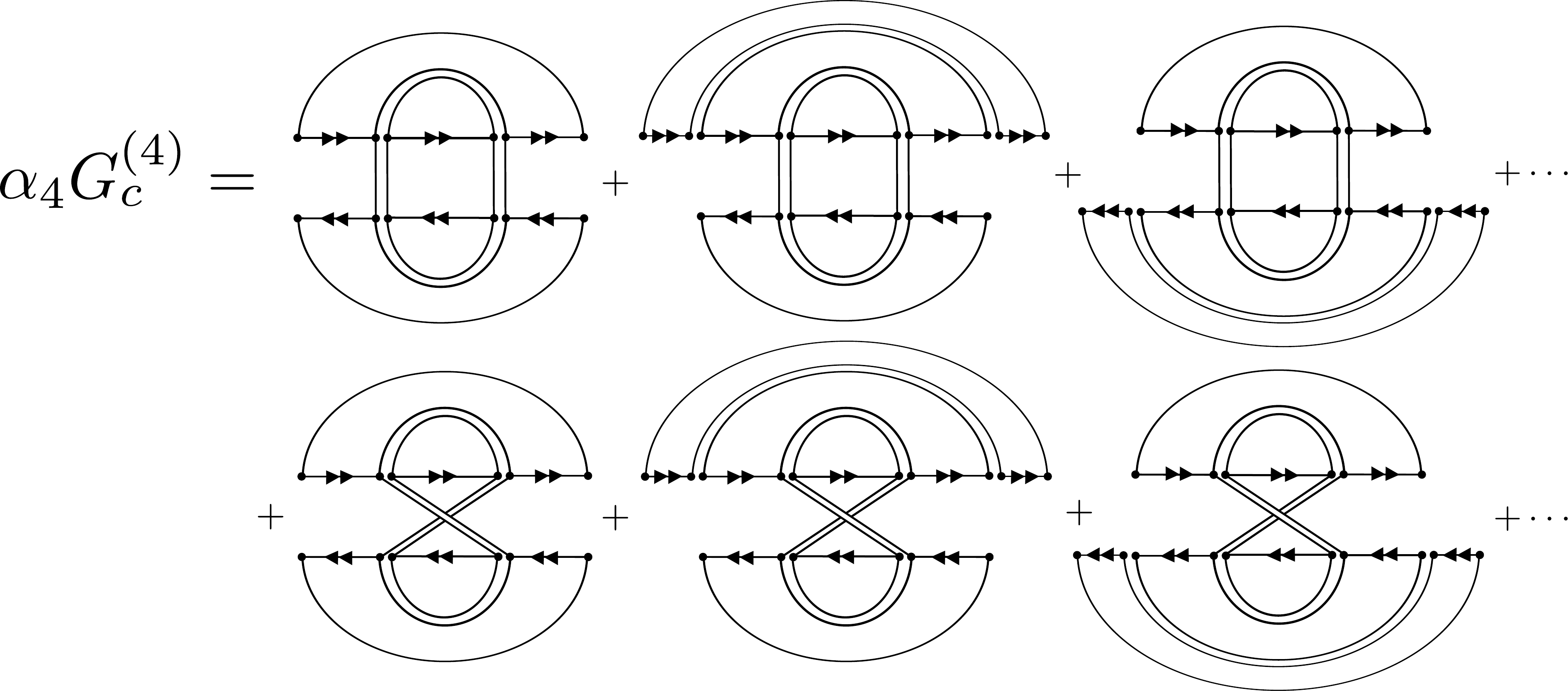}
	\label{fig:ladderdiagramssparse}
\end{figure}
where we have noted that it is possible to put the full series of rainbow diagrams in place of each propagator (as usual), and it is also possible place additional arcs over the central ribbon part without affecting the order in $N$. Importantly, we see that while diagrams that have crossing arcs (starting and finishing on the same horizontal line of propagators) give rise to subleading corrections, we see here that it is possible for vertical double lines to cross and not give rise to a subleading term. We saw the same thing occur in the GUE case -- cyclic permutations of the pairs of dynamic variables sharing the same time do not change the topology of the diagram. In the end, the above series evaluates to 
\begin{align}
	\alpha_4G^{(4)}_c(\omega,\mu) = \frac{1}{N}\frac{2\alpha_4 \sigma^4G^3(\omega)G^3(\mu)}{[1- \sigma^2 G^2(\omega)][1- \sigma^2 G^2(\mu)]} .
\end{align}
The diagrams that arise from the $S_\mathrm{het}$ contribution to the action are similar. However, in this case, there are four non-vanishing classes of diagram, as opposed to the two associated to $S_\mathrm{4}$
\begin{figure}[H]
	\centering 
	\includegraphics[scale = 0.15]{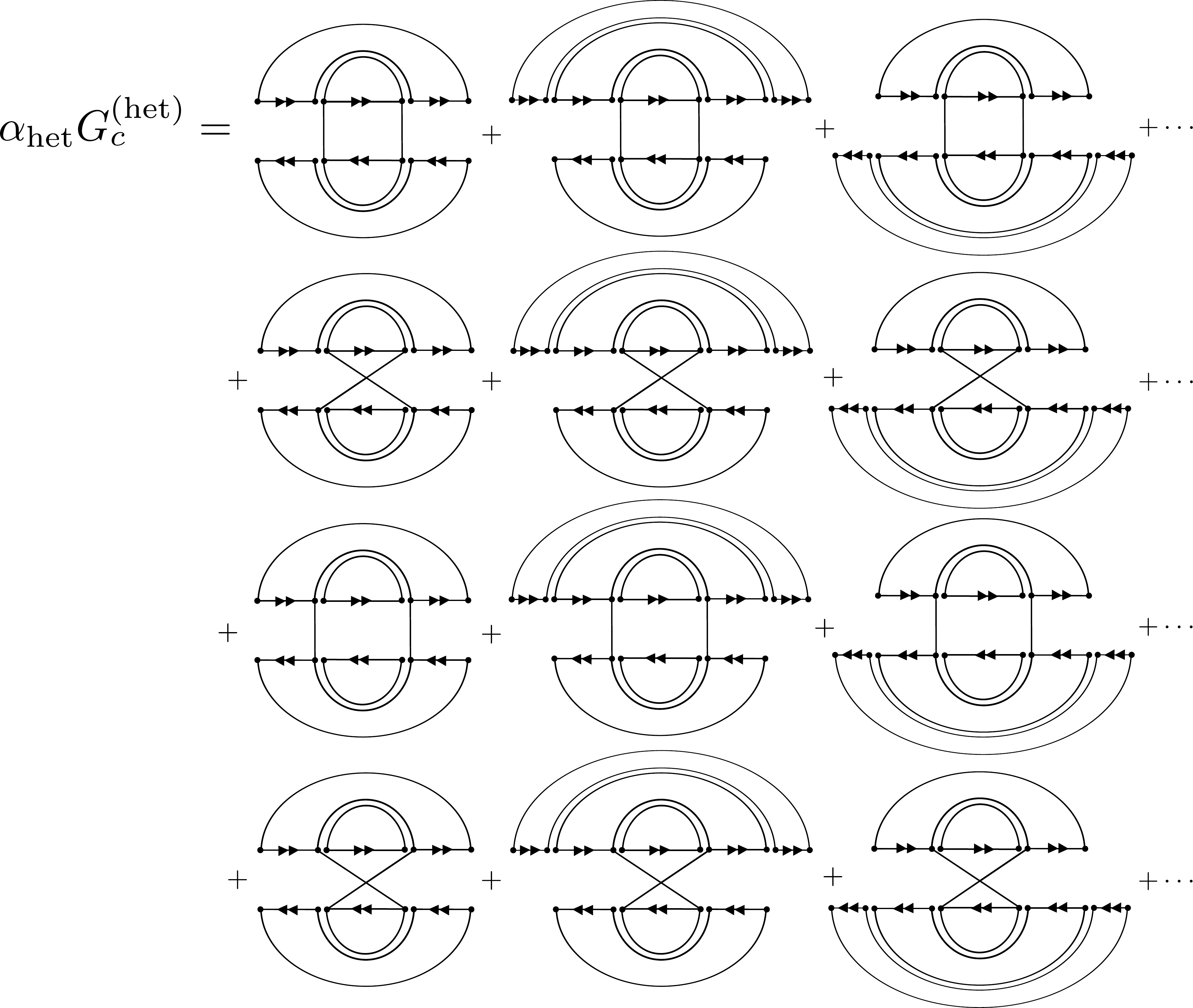}
	\label{fig:ladderdiagramsnetwork}
\end{figure}
In this case, the subseries evaluates to 
\begin{align}
	\alpha_\mathrm{het}G^{(\mathrm{het})}_c(\omega,\mu) = \frac{1}{N}\frac{4\alpha_\mathrm{het} \sigma^4G^3(\omega)G^3(\mu)}{[1- \sigma^2 G^2(\omega)][1- \sigma^2 G^2(\mu)]}.
\end{align}
Finally, let us consider the diagrams corresponding to $S_\mathrm{cyc}$. In this case, we find the following first-order contribution
\begin{figure}[H]
	\centering 
	\includegraphics[scale = 0.2]{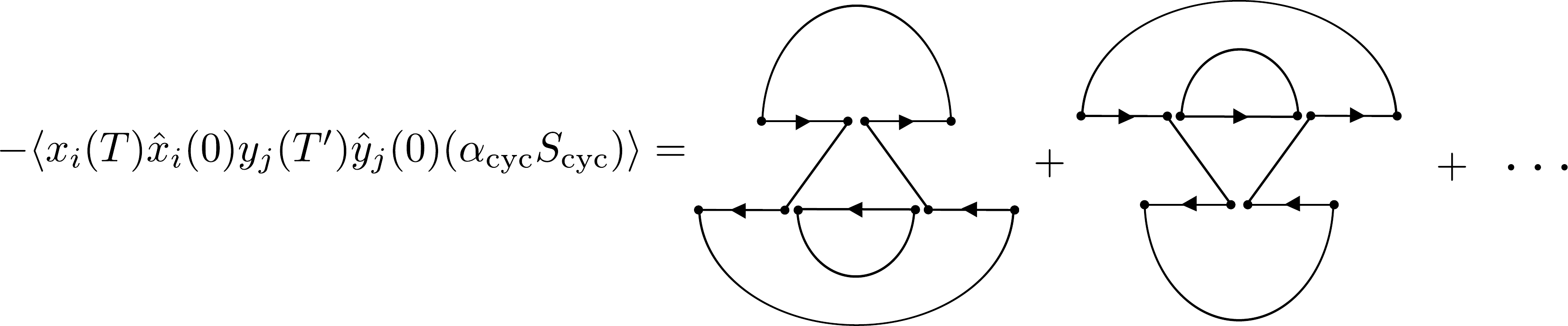}
	\label{fig:cyclicladder}
\end{figure}	
Unlike the $S_4$ and $S_\mathrm{het}$ cases, we see that the diagrams here simply go as $1/N^2$ as they did in the GOE case. Taking only the $O(1/N)$ contributions to $G_c$ as significant, we can therefore ignore the cyclic correlations. Ultimately, we obtain
\begin{align}
	G_c(\omega, \mu) &= G_c^{(\mathrm{GOE})} + \alpha_4 G_c^{(4)} +\alpha_\mathrm{het} G_c^{(\mathrm{het})} + \alpha_\mathrm{cyc} G_c^{(\mathrm{cyc})}+ O\left(\frac{\alpha^2}{N}\right) \nonumber \\
	&= \frac{1}{N}\frac{(2\alpha_4 +4\alpha_\mathrm{het}) \sigma^4G^3(\omega)G^3(\mu)}{[1- \sigma^2 G^2(\omega)][1- \sigma^2 G^2(\mu)]} + O\left(\frac{1}{N^2}\right) + O\left(\frac{\alpha^2}{N}\right). 
\end{align}
Using the expression for $G$ in Eq.~(\ref{gpluscorrection}), which accounts for the correction to the spectral edge, and using the formula in Eq.~(\ref{corrfromgc}), one arrives at the formula for $\rho_c(\omega, \mu)$ in Eq.~(7) of the main text. The function $f(\cdot)$ is given by 
\begin{align}
	f(\omega) = \frac{4\omega^2 (\bar \lambda^2-4) + 4 \omega \bar \lambda({\lambda'}^{2}+2) + {\lambda'}^2(4 + {\lambda'}^2)}{4\omega^2 (\bar \lambda^2-4) + 4 \omega \bar \lambda({\lambda'}^{2}+4) + (4 + {\lambda'}^2)^2 } \sqrt{(\lambda_+ - \omega)(\omega-\lambda_-)},
\end{align}
where $\lambda_- < \omega,\mu<\lambda_+$ and we use the shorthand $\bar \lambda = (\lambda_+ + \lambda_-)/2$, $\lambda' = \sqrt{\vert\lambda_+\lambda_-\vert} $, $\overline{\alpha}_2 = 2 \alpha_4 + 4 \alpha_\mathrm{het}$ and $\lambda_\pm$ are as given in Eq.~(\ref{gpluscorrection}). We verify this explicitly in Fig. \ref{fig:corrandden}.

\begin{figure}[h]
	\centering 
	\includegraphics[scale = 0.45]{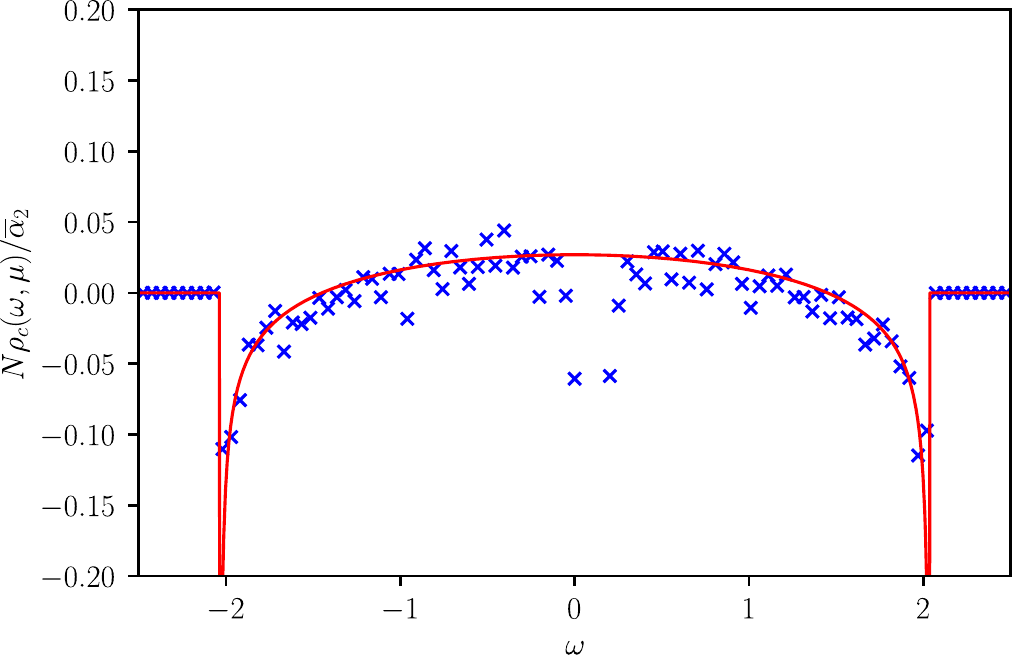}
	\captionsetup{justification=raggedright,singlelinecheck=false, font=small}
	\caption{Comparison of Eq.~(7) of the main text with the results of numerical diagonalisation of the signed adjacency matrices of Erd\H{o}s-R\'enyi graphs where $a_{ij} = a_{ji} = \pm 1/\sqrt{p}$ with probability $p/N$ and zero otherwise, giving $\sigma = 1$, $\alpha_4 = 1/p$ and $\alpha_\mathrm{het} = \alpha_\mathrm{cyc} = 0$. Here, $N = 2000$, $p = 30$ and $\mu = 0.1$ (where the $1/(\omega-\mu)^2$ divergence is observed). }\label{fig:corrandden}
\end{figure}

\subsection{$H_c$ and the local density of states fluctuations $K_c$}
We can also calculate other 2-point quantities using the diagrammatic method. In particular, to calculate $H_c(\omega, \mu)$ and thus extract $K_c(\omega, \mu)$ [see Eqs.~(\ref{hcdef}) and (\ref{corrfromkc})], we wish to evaluate the following series diagrammatically to first order in $\alpha$
\begin{align}
	&\frac{1}{N}\sum_{k} \left\langle R^{(x)}_{kk}(T,0) R^{(y)}_{kk}(T',0) \right \rangle - \frac{1}{N^2}\sum_{k,l} \left\langle R^{(x)}_{kk}(T,0) \right\rangle \left\langle R^{(y)}_{ll}(T',0) \right \rangle    \nonumber \\
	&= \frac{-1}{N}\sum_{k}\sum_{r, r_1, r_2, r_3} \left\langle \frac{(S_\mathrm{GOE})^r (\alpha_4 S_\mathrm{4})^{r_1} (\alpha_\mathrm{het}S_\mathrm{het})^{r_2} (\alpha _\mathrm{cyc}S_\mathrm{cyc})^{r_3}}{r!\, r_1! \,r_2!\, r_3!} x_k(T) \hat x_k(0)  y_k(T') \hat y_k(0) \right\rangle_0 \nonumber \\
	&+\frac{1}{N^2}\sum_{k, l}\sum_{\substack{r, r_1, r_2, r_3 \\ r', r_1', r_2', r_3'}} \left\langle \frac{(S_\mathrm{GOE})^r (\alpha_4 S_\mathrm{4})^{r_1} (\alpha_\mathrm{het}S_\mathrm{het})^{r_2} (\alpha _\mathrm{cyc}S_\mathrm{cyc})^{r_3}}{r!\, r_1! \,r_2!\, r_3!} x_k(T) \hat x_k(0)\right\rangle_0 \nonumber \\
	&\hspace{2cm}\times\left\langle \frac{(S_\mathrm{GOE})^{r'} (\alpha_4 S_\mathrm{4})^{r_1'} (\alpha_\mathrm{het}S_\mathrm{het})^{r_2'} (\alpha _\mathrm{cyc}S_\mathrm{cyc})^{r_3'}}{r'!\, r_1'! \,r_2'!\, r_3'!} y_l(T') \hat y_l(0) \right\rangle_0.\label{serieskc}
\end{align}
In this case, the series of diagrams is even simpler than in the calculation of $G_c$ above. Again, only contributions from $S_4$ and $S_\mathrm{het}$ are non-vanishing, while those from $S_\mathrm{cyc}$ vanish. However, this time, the only non-vanishing diagrams at first order in $\alpha$ are 
\begin{figure}[H]
	\centering 
	\includegraphics[scale = 0.18]{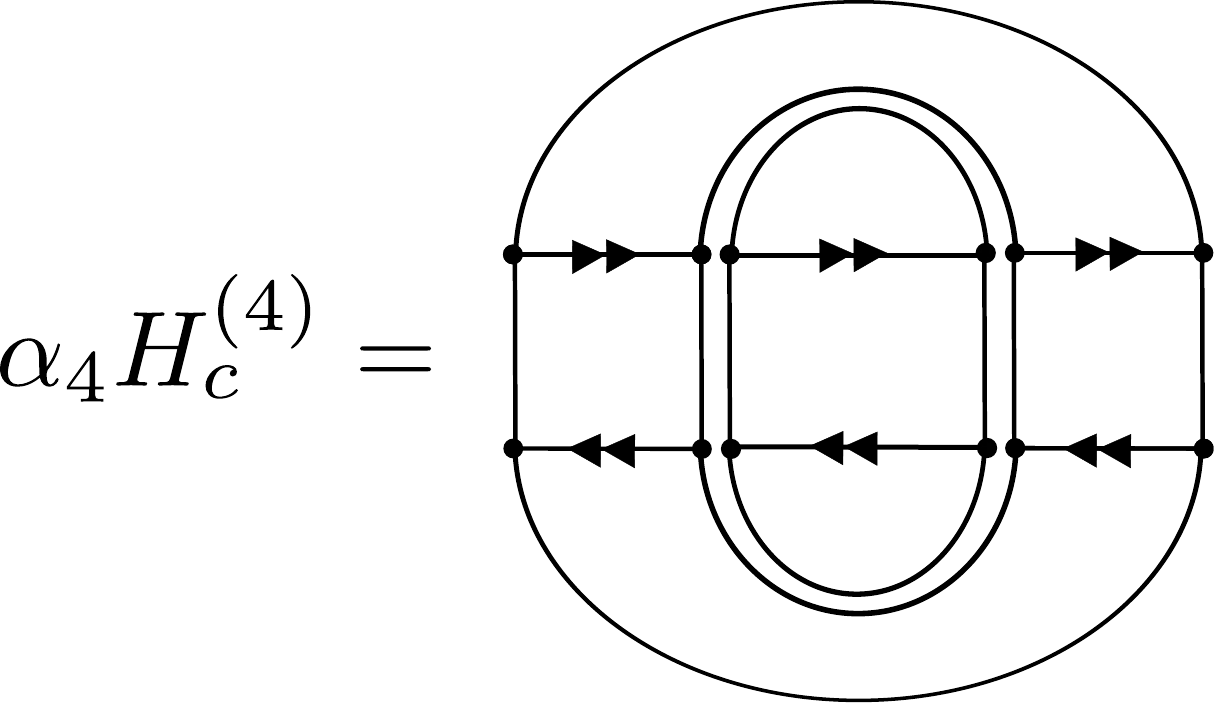}\hspace{3cm}
	\includegraphics[scale = 0.18]{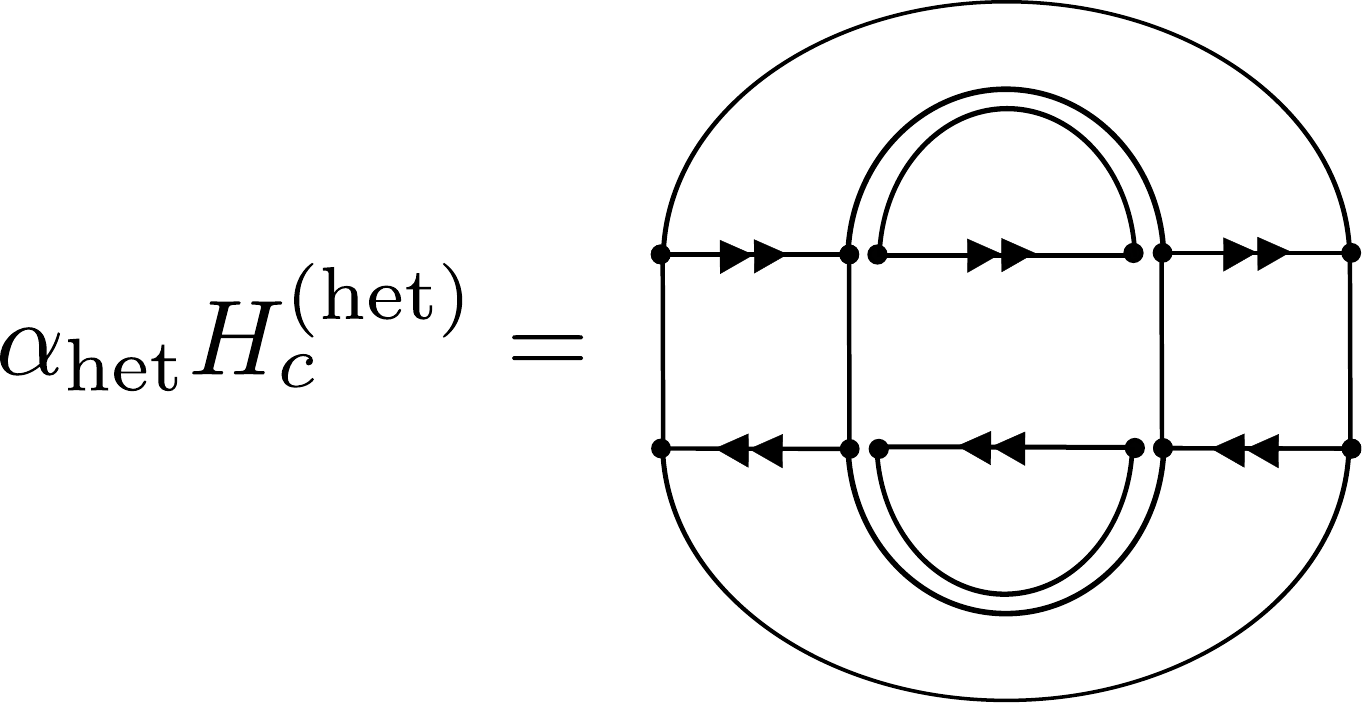}
	\label{fig:ladderdiagramskc}
\end{figure}
We see here that, due to the factor of $N^{-1}$ in Eq.~(\ref{serieskc}) that replaces the factor of $N^{-2}$ in Eq.~(\ref{expsum2pointmod}), these diagrams are of the order $N^0$. Just as one found that $G_c$ vanished in the GOE but remained finite in the non-GOE cases, one therefore finds that $K_c \sim N^{-1}$ in the GOE case, whereas $K_{c} \sim N^0$ when $\alpha_4 \neq 0$ or $\alpha_\mathrm{het} \neq 0$. 

The diagrammatic series thus evaluates to
\begin{align}
	H_c(\omega,\mu) = (\alpha_4+ \alpha_\mathrm{het}) \sigma^4 G^3(\omega)G^3(\mu)+ O\left(\frac{1}{N}\right) + O(\alpha^2). \label{hctheory}
\end{align}
We therefore arrive at the compact expression [using Eq.~(\ref{corrfromkc})]
\begin{align}
	K_c(\omega, \mu) = \frac{\overline{\alpha}_3}{4 \pi^2}c(\omega)c(\mu) + O\left(\frac{1}{N}\right) + O(\alpha^2) , \label{kctheory}
\end{align}
where in this case $c(\omega) = \frac{1}{4}[4 \omega^2 - \omega(\lambda_+ + \lambda_-)+ \lambda_-\lambda_+ ] \sqrt{(\lambda_+ - \omega)(\omega-\lambda_-)} $, and we have $\overline{\alpha}_3 = \alpha_4 + \alpha_\mathrm{het}$. This is tested below in Fig. \ref{fig:kc}.

\begin{figure}[H]
	\centering 
	\includegraphics[scale = 0.48]{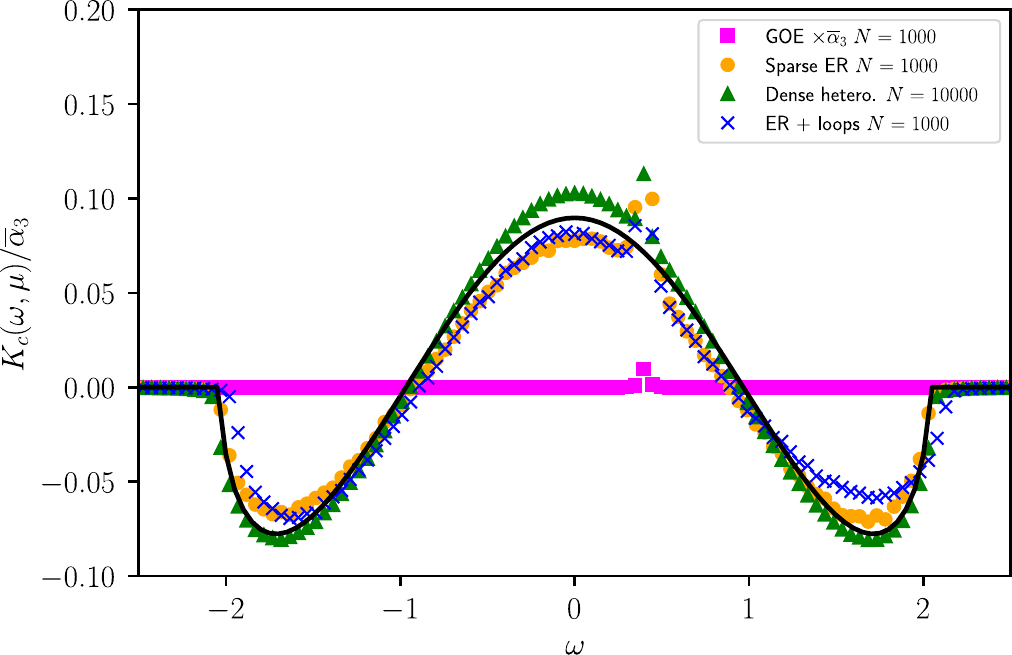}
	\captionsetup{justification=raggedright,singlelinecheck=false, font=small}
	\caption{Test of the expression for the covariance of the local density of states fluctuations in Eq.~(\ref{kctheory}) using the same ensembles used in Fig. 2 of the main text (see Section \ref{subsection:ensembles} for details).}\label{fig:kc}
\end{figure}

\section{Heterogeneous mean-field theory}
We now comment on how the non-zero value of $K_c$ (or equivalently $\alpha_4$ and/or $\alpha_\mathrm{het}$) indicates the necessity of a heterogeneous mean-field theory (as opposed to a homogeneous mean-field theory), and how $\rho_c$, which was available to us from the diagrammatic approach, is not captured by the heterogeneous mean-field theory.

Let us consider once again the dynamical system in Eq.~(\ref{dynamicalsystem}). In Dynamical Mean-Field Theory (DMFT), one attempts to replace the original dynamics in Eq.~(\ref{dynamicalsystem}) with a simpler `effective dynamics', which is valid in the limit $N\to \infty$ and replicates the statistical behaviour of the original system. The effective dynamics that we seek decouples the individual components $x_i$, replacing the interactions with noise and memory terms. In a homogeneous mean-field description, each component is statistically identical. We demonstrate here that non-Gaussianity leads to non-negligible statistical heterogeneities that survive even in the limit $N \to \infty$, leading to the necessity of a heterogeneous mean-field theory. 

We consider once again the MSRJD action corresponding the process in Eq.~(\ref{dynamicalsystem}). Ignoring the $y$-type variables, which are no longer helpful, we have again
\begin{align}
	S \approx S_0 + S_\mathrm{GOE} + \alpha_4 S_\mathrm{4} + \alpha_\mathrm{het} S_\mathrm{het} + \alpha_\mathrm{cyc} S_\mathrm{cyc} + O(\alpha^2), 
\end{align}
but now
\begin{align}
	S_0 &=  i\sum_{i} \int dt \, \hat x_i (\dot x_i  + \omega x_i),\nonumber \\
	S_\mathrm{GOE} &=   -\frac{\sigma^2}{2\times2! N} \sum_{ij} \int dt_1dt_2 [\hat x_i(t_1) x_j(t_1) + \hat x_j(t_1) x_i(t_1) ] [\hat x_i(t_2) x_j(t_2) + \hat x_j(t_2) x_i(t_2)  ] , \nonumber \\
	S_\mathrm{4} &=  \frac{\sigma^4}{2 \times 4!\, N} \sum_{i,j} \int dt_1 \cdots dt_4 [\hat x_i(t_1) x_j(t_1) + \hat x_j(t_1) x_i(t_1) ]\times\cdots\times[\hat x_i(t_4) x_j(t_4) + \hat x_j(t_4) x_i(t_4)  ]  , \nonumber \\
	S_\mathrm{het} &= \frac{\sigma^4}{ 2\times(2!)^2 N^2}  \sum_{i,j,k} \int dt_1 \cdots dt_4 [\hat x_i(t_1) x_j(t_1) + \hat x_j(t_1) x_i(t_1) ][\hat x_i(t_2) x_j(t_2) + \hat x_j(t_2) x_i(t_2) ]\nonumber\\
	&\hspace{5cm}\times[\hat x_i(t_3) x_k(t_3) + \hat x_k(t_3) x_i(t_3)  ][\hat x_i(t_4) x_k(t_4) + \hat x_k(t_4) x_i(t_4)  ]  \nonumber \\
	S_\mathrm{cyc} &= i\frac{\sigma^3}{3! \, N^2} \sum_{i,j,k} \int dt_1 dt_2dt_3  [\hat x_i(t_1) x_j(t_1) + \hat x_j(t_1) x_i(t_1) ] [\hat x_j(t_2) x_k(t_2) + \hat x_k(t_2) x_j(t_2) ] [\hat x_k(t_3) x_i(t_3) + \hat x_i(t_3) x_k(t_3) ].
\end{align}
Following, for example, Refs. \cite{baron2022eigenvalues, galla2024generating}, we now wish to introduce `order parameters', which will allow the factorisation of the MSRJD path integral. We define
\begin{align}
	C(t_1,t_2) = \frac{1}{N}\sum_i x_i(t_1) x_i(t_2), \,\,\,\,\,
	K(t_1,t_2) &= \frac{1}{N}\sum_i x_i(t_1) \hat x_i(t_2), \,\,\,\,\,
	L(t_1,t_2) = \frac{1}{N}\sum_i \hat x_i(t_1) \hat x_i(t_2), \nonumber \\
	C_{4}(t_1,t_2,t_3,t_4) = \frac{1}{N}\sum_i x_i(t_1) x_i(t_2) x_i(t_3) x_i(t_4), &\,\,\,\,\,
	L_4(t_1,t_2, t_3, t_4) = \frac{1}{N}\sum_i \hat x_i(t_1) \hat x_i(t_2) \hat x_i(t_3) \hat x_i(t_4), \nonumber \\
	K_{2,2}(t_1,t_2;t_3,t_4) &= \frac{1}{N}\sum_i x_i(t_1) x_i(t_2)\hat x_i(t_3) \hat x_i(t_4), \,\,\,\,\, \nonumber \\
	K_{3,1}(t_1,t_2,t_3;t_4) = \frac{1}{N}\sum_i x_i(t_1) x_i(t_2) x_i(t_3) \hat x_i(t_4), &\,\,\,\,\,
	K_{1,3}(t_1;t_2, t_3, t_4) = \frac{1}{N}\sum_i x_i(t_1) \hat x_i(t_2) \hat x_i(t_3) \hat x_i(t_4).
\end{align}
Performing the usual saddle-point procedure \cite{galla2024generating}, and using that $\langle L(t,t')\rangle = \langle K_{1,3}(t_1;t_2, t_3, t_4) \rangle =\langle K_{3,1}(t_1,t_2,t_3;t_4)\rangle = \langle L_4(t_1,t_2, t_3, t_4)\rangle = 0$, $\langle K_{2,2}(t_1,t_2;t_3,t_4)\rangle = \langle K(t_1,t_3)\rangle \langle K(t_2,t_4)\rangle + \langle K(t_1,t_4)\rangle \langle K(t_2,t_3)\rangle + O(\alpha) $ and  $\langle C_4(t_1,t_2;t_3,t_4)\rangle = \langle C(t_1,t_3)\rangle \langle C(t_2,t_4)\rangle + \langle C(t_1,t_4)\rangle \langle C(t_2,t_3)\rangle+ \langle C(t_1,t_2)\rangle \langle C(t_3,t_4)\rangle + O(\alpha) $, one succeeds in factorising the action, which then represents $N$ independent `effective' processes. The effective process can be written (accurate up to leading order in $\alpha$)
\begin{align}
	\dot x_z =& - \omega x_z + \sigma^2 (1 + z)\int dt' R(t,t') x_z(t') + \alpha_\mathrm{het} \sigma^4 \int dt_1 dt_2 dt_3 R(t, t_1) R(t_1,t_2) R(t_2,t_3) x_z(t_3) \nonumber \\
	&+ \alpha_\mathrm{cyc} \sigma^3 \int dt_1 dt_2 R(t,t_1)R(t_1, t_2)x_z(t_2)  + \xi(t), \label{effectiveprocess}
\end{align}
where $R(t,T) = \langle\delta x(t)/\delta \xi(T) \vert_{\xi = 0}\rangle$ are the disorder-averaged response functions, $z$ is a quenched Gaussian random variable with variance
\begin{align}
	\langle z^2 \rangle = \alpha_4 + \alpha_\mathrm{het} ,
\end{align}
and $\xi(t)$ are time-varying correlated non-Gaussian random variables with a self-consistent correlator, which we do not give here for the sake of brevity. The precise form of the noise $\xi$ is not relevant for the present discussion.

We see that in the limit $\alpha_4, \alpha_\mathrm{het} \to 0$, we can effectively set $z = 0$, and we recover the result for the effective process in Ref. \cite{baron2022eigenvalues}. In this case, the components of the system become statistically homogeneous, and the system's behaviour is thus described by a homogeneous mean-field theory. With non-zero $\alpha_4$ and $\alpha_\mathrm{het}$, we instead have a heterogeneous mean-field theory, where we not only have to average over the self-consistent noise, which affects each $x_i$ in a statistically equivalent manner; we also must average over a quenched randomness which varies across sites $i$. 

Let us now attempt to find the resolvent $G(\omega) = \mathcal{L}_t[R(t,0)](\eta)$ from the effective process Eq.~(\ref{effectiveprocess}). Functionally differentiating, taking the Laplace transform and rearranging, we have
\begin{align}
	G_z(\omega) = \left[\omega - \sigma^2 (1+ z)G - \alpha_\mathrm{het} \sigma^4 G^3 - \sigma^3 \alpha_\mathrm{cyc} G^2\right]^{-1}.
\end{align}
Expanding this expression to $O(\alpha)$, we obtain
\begin{align}
	G_z(\omega) = \frac{1}{\omega - \sigma^2 G} + z \sigma^2 G^3 + \sigma^4 z^2 G^5 + \alpha_\mathrm{het}\sigma^4 G^5 + \alpha_\mathrm{cyc} \sigma^3 G^4 . \label{gz}
\end{align}
Averaging this expression over $z$, one obtains the expression for $G$ in Eq.~(\ref{gselfcons}), and thus recovers the modified semi-circle law of Eq.~(\ref{modsemigeneral}). 

We can also understand the origin of the on-site variation $K_c$ using the expression in Eq.~(\ref{gz}). Ignoring terms of the order $O(\alpha^2)$, one finds simply
\begin{align}
	G_z(\omega) - G(\omega) =  z \sigma^2 G^3 ,
\end{align}
from which the expression in Eq.~(\ref{hctheory}), and consequently Eq.~(\ref{kctheory}), follows using $H_c(\omega, \mu) = \langle [G_z(\omega) - G(\omega)][G_z(\mu) - G(\mu)]\rangle_z$.

However, we see that the heterogeneous mean-field theory is entirely inadequate for recovering the two-point functions that were the focus of the main text. The site-to-site correlation, quantified by $G_c$ and $\rho_c$, is a subleading effect in $1/N$, which is neglected during the saddle-point procedure in which we factorise the MSRJD functional integral. In order to recover the results for the 2-point function $G_c$, we would no longer be able to treat the components of the system as independent. We would have to reinstate the site index in the effective process, and take into account correlations between the noise $\xi_i(t)$ and the quenched variables $z_i$ at different sites $i$. This failure is to be expected from a mean-field approach (even the heterogeneous mean-field theory), which ignores correlations by definition.

\section{Power-law-tailed eigenvector statistics}
\begin{figure}[H]
	\centering 
	\includegraphics[scale = 0.48]{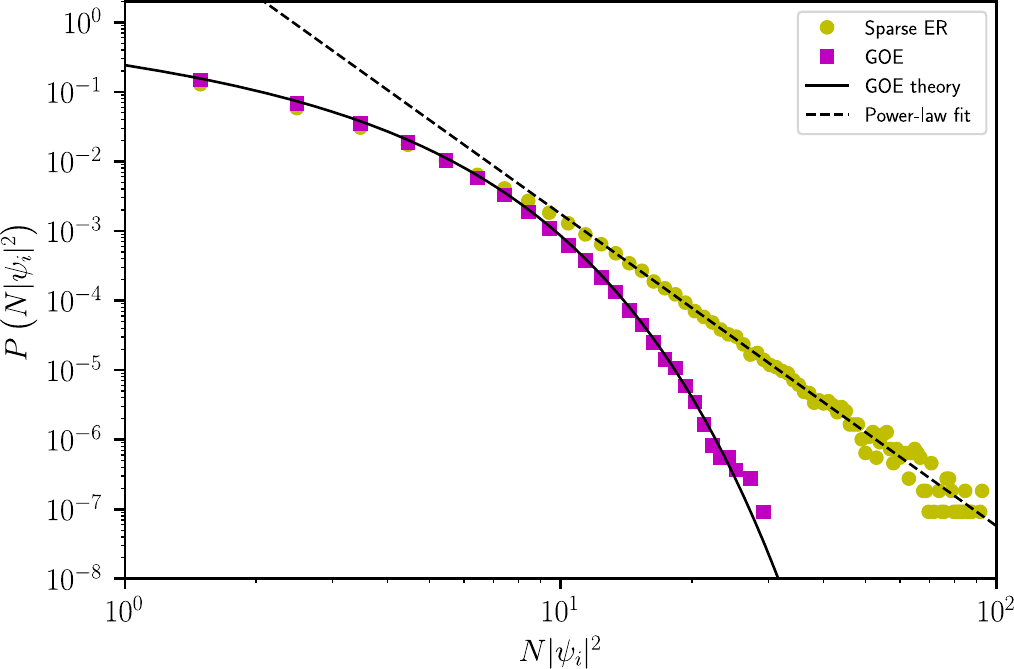}
	\includegraphics[scale = 0.48]{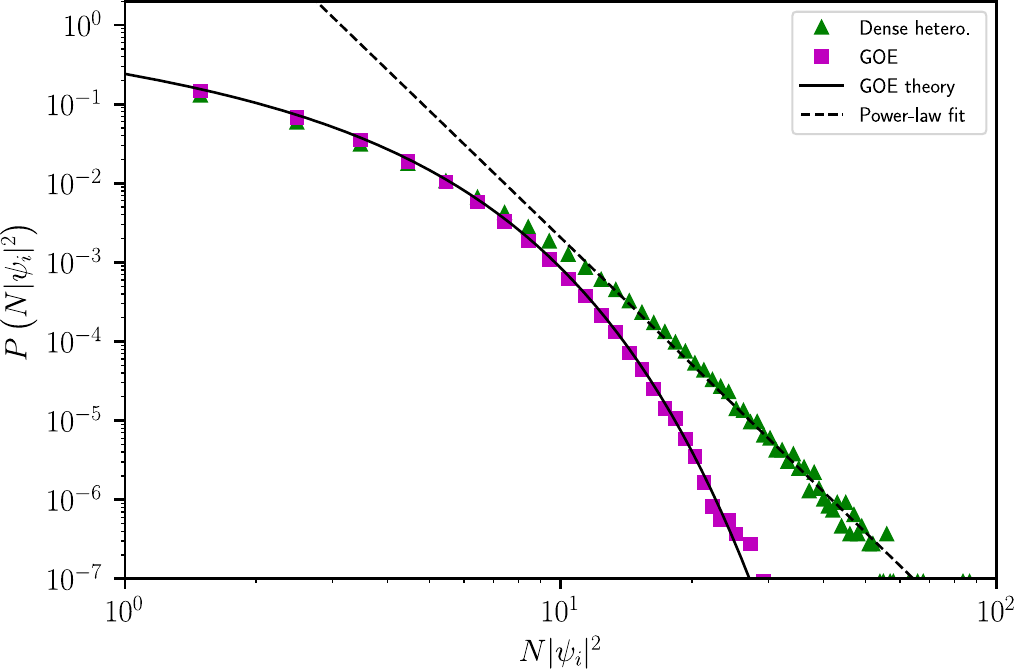}
	\captionsetup{justification=raggedright,singlelinecheck=false, font=small}
	\caption{Similar to Fig. 3 in the main text. We once again see the clear emergence of power-law tails, which are fitted very well by a straight line on the log-log axes. (Left) Signed ER graph with $p = 5$, defined in Eq.~(\ref{sparsedef}). (Right) Chung-Lu model with a uniform degree distribution, with $s^2 = 0.25$ defined in Eq.~(\ref{ssqdef}). In both cases, $N = 10000$ and results are averaged over 1000 realisations. One sees that the slope of the fitted line (i.e. the power-law exponent) varies depending on the ensemble and the values of $\alpha_4$ and $\alpha_\mathrm{het}$.}\label{fig:powerlawtails}
\end{figure}